\documentclass[12pt]{article}
\usepackage{a4wide}
\usepackage{amsmath,amsthm,amsfonts}
\usepackage{graphicx,epsfig,array}

\newtheorem{theorem}{Theorem}[section]

\newtheorem{remark}[theorem]{Remark}

\newcommand{\url}[1]{{\tt \small #1}}
\newcommand{\Ex}[2]{\mathbb{E}_{#1}\!\left[\,#2\,\right]}
\newcommand{\Qx}[1]{\mathbb{Q}\left\{\,#1\,\right\}}

\newcommand{\defleg}{\mbox{D{\tiny EF}L{\tiny EG}}}
\newcommand{\prmleg}{\mbox{P{\tiny REMIUM}L{\tiny EG}}}
\newcommand{\dv}{\mbox{D\tiny{V01}}}
\newcommand{\npv}{\mbox{N\tiny{PV}}}

\newcommand{\gpl}{Z}

\newcommand{\defrate}{C}
\newcommand{\defratenorm}{{\bar\defrate}}
\newcommand{\loss}{L}
\newcommand{\lossnorm}{{\bar\loss}}
\newcommand{\losstr}{\lossnorm^{A,B}}
\newcommand{\cds}{S}
\newcommand{\cdstr}{S^{A,B}}
\newcommand{\upfronttr}{U^{A,B}}
\newcommand{\ind}[1]{1_{\{#1\}}}
\newcommand{\rec}{R}

\def\onehalf{\tfrac{1}{2}}

\title{\vspace{-1cm}
{\normalsize A vastly extended and updated version of this paper will appear as a book:}\\
{\normalsize \bf Credit Models and the Crisis: A journey into CDOs, Copulas,
Correlations and Dynamic Models, Wiley, Chichester, 2010
}\\ --- \\ {\Large \bf Credit Models and the Crisis, or:\\  How I learned to stop worrying and love the CDOs}
}
\author{
Damiano Brigo \ \ \
Andrea Pallavicini \ \ \
Roberto Torresetti\thanks{Brigo is with the Dept. of Mathematics, Imperial College, {\tt d.brigo@imperial.ac.uk}; Pallavicini is with Banca Leonardo, {\tt andrea.pallavicini@bancaleonardo.com};  Torresetti is with BBVA, {\tt roberto.torresetti@grupobbva.com}. This paper reflects the authors opinion and not necessarily those of their employers. First Version: March 17, 2009. First Posted on SSRN.com and arXiv.org on Dec. 29, 2009.  This version: \today
}
}
\date{}

\begin{document}

\maketitle

\begin{abstract}

We follow a long path for Credit Derivatives and Collateralized Debt Obligations (CDOs) in particular, from the introduction of the Gaussian copula model and the related implied correlations to the introduction of arbitrage-free dynamic loss models capable of calibrating all the tranches for all the maturities at the same time. En passant, we also illustrate the implied copula, a method that can consistently account for CDOs with different attachment and detachment points but not for different maturities. The discussion is abundantly supported by market examples through history. The dangers and critics we present to the use of the Gaussian copula and of implied correlation had all been published by us, among others, in 2006, showing that the quantitative community was aware of the model limitations before the crisis. We also explain why the Gaussian copula model is still used in its base correlation formulation, although under some possible extensions such as random recovery. Overall we conclude that the modeling effort in this area of the derivatives market is unfinished, partly for the lack of an operationally attractive single-name consistent dynamic loss model, and partly because of the diminished investment in this research area.

\end{abstract}

{\small \bf JEL classification code: C15, C31, C46, C61, G12, G13. \\ \indent AMS classification codes: 60J75, 62E20, 91B30, 91B70, 91B82}

\medskip



\section*{Preface}
\addcontentsline{toc}{section}{Preface}
\subsection*{Part 1: Randomized by foolishness}
\addcontentsline{toc}{subsection}{Part 1: Randomized by foolishness}
This is a paper that has been written very rapidly in order to respond to a perceived need of clarity in the quantitative world. The paper subtitle,  ``How I learned to stop worrying and love the CDOs" is obviously ironic in referring to the hysteria that has often characterized accounts of modeling and mathematical finance in part of the press and the media, and the demonization of part of the market products related to the crisis, such as CDOs and derivatives more generally. Let us be clear and avoid any misunderstanding: The crisis is very real, it has caused suffering to many individuals, families and companies. However, it does not help looking for a scapegoat without looking at the whole picture with a critical eye. Accounts that try and convince the public that the whole crisis is due mainly to modeling and to sophisticated and obscure products being traded are necessarily partial, and to partly righting this perceptive bias this paper is devoted.
%

Indeed, the public opinion has been bombarded with so many cliches on derivatives, modeling and quantitative analysis that we feel a paper offering a little clarity is needed. And while we are aware that this sounds a little Don Quixotesque, we hope this paper will help changing the situation. In trying to do so, we need to balance carefully the perspectives of different readerships. We would like our paper to be attractive to a relatively general industry and academic public without disappointing the scientific and technically minded specialists, and at the same time we do not want our  paper to be a best-seller type of publication full of bashing and negatively provocative ideas and very little actual technical content. All of this while keeping windmills at large\footnote{The Ingenious Hidalgo Don Quixote of La Mancha, 1605 and 1615.}. Hence we will be walking the razor's hedge here, in trying to maintain the balance between a popular account and scientific discourse.

We are not alone in our attempt to bring clarity\footnote{See primarily Shreve, S. (2008), Don't Blame the Quants, Forbes Commentary, but also, for example, Donnelly, C., and Embrechts, P. (2009), The devil is in the tails: actuarial mathematics and the subprime mortgage crisis, accepted for publication in the ASTIN Bulletin. Giorgio  Szego (2009), the Crash Sonata in D Major (to appear in the Journal of Risk Management in Financial Institutions), gives a much broader overview of the crisis with some critical insight that is helpful in clarifying the more common misconceptions}. This paper however does so in an extensive technical way, showing past and present research that is quite relevant in disproving a number of mis-conceptions on the role of mathematics and quantitative analysis in relation with the crisis. This paper takes an extensive technical path, starting with static copulas and ending up with dynamic loss models. Even though our paper is short, we follow a long path for Credit Derivatives and multi-name credit derivatives in particular, focusing on Collateralized Debt Obligations (CDOs). What are CDOs?\index{CDO}\index{Collateralized Debt Obligation|see{CDO}} To describe the simplest possible CDO, say a synthetic CDO on the corporate market, we can proceed as follows.

We are given a portfolio of names, say 125 names for example. The names may default, generating losses to investors exposed to those names. In a CDO tranche\index{CDO!tranche} there are two parties, a protection buyer and a protection seller. A tranche is a portion of the loss of the portfolio between two percentages. For example, the 3-6\% tranche focuses on the losses between 3\% (attachment point) and 6\% (detachment point). Roughly speaking, the protection seller  agrees to pay to the buyer all notional default losses (minus the recoveries) in the portfolio whenever they occur due to one or more defaults of the entities, within 3\% and 6\% of the total pool loss. In exchange for this, the buyer pays the seller a periodic fee on the notional given by the portion of the tranche that is still ``alive" in each relevant period.

In a sense, CDOs look like contracts selling (or buying) insurance on portions of the loss of a portfolio. The valuation problem is trying and determining the fair price of this insurance.

The crucial observation here is that ``tranching" is a non-linear operation. When computing the price (mark to market) of a tranche at a point in time, one has to take the expectation of the future tranche losses under the pricing measure. Since the tranche is a nonlinear function of the loss, the expectation will depend on all moments of the loss and not just on the expected loss. If we look at the single names in the portfolio, the loss distribution of the portfolio is characterized by the marginal distributions of the single names defaults and by the dependency among different names' defaults. Dependency is commonly called, with an abuse of language, ``correlation". This is an abuse of language because correlation is a complete description of dependence for jointly Gaussian random variables, but more generally it is not. The complete description is either the whole multivariate distribution or the so-called ``copula function", that is the multivariate distribution once the marginal distributions  have been standardized to uniform distributions.

The dependence of the tranche on ``correlation" is crucial. What the market does is assuming a Gaussian Copula\index{Gaussian Copula} connecting the defaults of the 125 names. This copula is parametrized by a matrix with 7750 entries of pairwise correlation parameters. However, when looking at a tranche these 7750 parameters are assumed to be all equal to each other. So one has a unique parameter. This is such a drastic simplification that we need to make sure it is noticed:

\[   7750\ \ \mbox{parameters} \longrightarrow \ \  1 \ \ \mbox{parameter.}\]

Then one chooses the tranches that are liquid on the market for standardized portfolios, for which the market price is known as these tranches are quoted. The unique correlation parameter is then reverse-engineered to reproduce the price of the liquid tranche under examination. This is called implied correlation, and once obtained it is used to value related products. The problem is that whenever the tranche is changed, this implied correlation changes as well. Therefore, if at a given time the 3-6\% tranche for a five year maturity has a given implied correlation, the 6-9\% tranche for the same maturity will have a different one. It follows that the two tranches on the same pool are priced with two models having different and inconsistent loss distributions, corresponding to the two different correlation values that have been implied.

This may sound negative, but as a matter of fact the situation is even worse. We will explain in detail that there are two possible implied correlation paradigms: compound correlation and base correlation. The second one is the one that is prevailing in the market. However, base correlation is inconsistent even at single tranche level, in that it prices  the 3-6\% tranche by decomposing it into the 0-3\% tranche and 0-6\% tranche and using two different correlations (and hence distributions) for those. Therefore base correlation is inconsistent already at single tranche level. And this inconsistency shows up occasionally in negative losses (i.e. in defaulted names resurrecting).

This is admittedly enough to spark a debate. Even before modeling enters the picture, some famous market protagonists have labeled the objects of modeling, i.e. derivatives, as responsible for a lot of troubles. Warren Buffett, in a very interesting 2003 report wrote: ``[...] Charlie and I are of one mind in how we feel about derivatives and the trading activities that go with them: We view them as time bombs, both for the parties that deal in them and the economic system. [...] In our view [...] derivatives are financial weapons of mass destruction, carrying dangers that, while now latent, are potentially lethal. [...]  The range of derivatives contracts is limited only by the imagination of man (or sometimes, so it seems, madmen)."\footnote{February, 21 2003, ``Berkshire Hathaway Inc. Annual Report 2002",\\ \url{www.berkshirehathaway.com/2002ar/2002ar.pdf}.}
While when hearing about products such as Constant Proportion Debt Obligations (CPDOs)\index{CPDO}\index{Constant Proportion Debt Obligation|see{CPDO}} or CDO squared\index{CDO!squared} one may sympathize with Mr. Buffett, this overgeneralization might be a little excessive. Derivatives, when used properly, can be quite useful. For example, swap contracts on several asset classes (interest rates, foreign exchange, oil and other commodities) and related options allow entities to trade risks and buy protection against adverse market moves. Without derivatives, companies could not protect themselves against adverse future movements of the prices of oil, exchange rates, interest rates etc. This is not to say that derivatives cannot be abused. They certainly can, and we invite the interested readers to reason on the case of CPDOs\footnote{See for example Torresetti and Pallavicini (2007), ``Stressing Rating Criteria Allowing for Default Clustering: the CPDO case``, and the Fitch Ratings report ``First Generation CPDO: Case Study on Performance and Ratings", published before the crisis in April 2007, stating ``[...] Fitch is of the opinion that the past 10 years by no means marked a high investment grade stress in the range of 'AAA' or 'AA'."
} as an example, and to read the whole report by Mr Buffett.\index{CPDO}

When moving beyond the products and entering the modeling issues, one may still find popular accounts resorting to quite colorful expressions such as ``the formula that killed Wall Street". Indeed, if one looks at popular accounts such as Salmon (2009)\footnote{Recipe for disaster: the Formula that killed Wall Street. Wired Magazine, 17.03.}, or Jones (2009)\footnote{Jones, S. (2009). Of couples and copulas: the formula that felled Wall St. April 24 2009, Financial Times.} just to make two examples, one may end up with the impression that the quantitative finance (``quant") community has been incredibly naive in accepting the Gaussian Copula\index{Gaussian Copula} and implied correlation without questioning it, possibly leading to what Mr Buffet calls ``mark to myth" in his above--mentioned report, especially when applying the calibrated correlation to other non-quoted ``bespoke tranches". In fact both articles have been written on the Gaussian copula, a static model that is little more than a static multivariate distribution which is used in credit derivatives (and in particular CDOs) valuation and risk management. Can this simple static model have fooled everyone in believing it was an accurate representation of a quite dynamic reality, and further cause the downfall of Wall Street banks? While Salmon (2009) correctly reports that some of the deficiencies of the model have been known for a while, Jones (2009) in the Financial Times wonders why no-one seemed to have noticed the model's weaknesses. The crisis is considered to have been heavily affected by mathematical models, with the accent on ``mathematical".

This is in line with more general criticism of anything quantitative appearing in the news. As an example, the news article ``McCormick Bad Dollars Derive From Deficits Model Beating Quants",\footnote{By Oliver Biggadike, November, 24 2009, Bloomberg.} more focused on the currency markets, informs us that ``[...] focus on the economic reasons for currency moves is gaining more traction after years when traders and investors relied on mathematical models of quantitative analysis." Then it continues with  ``These tools worked during times of global growth and declining volatility earlier this decade, yet failed to signal danger before the financial crisis sparked the biggest currency swings in more than 15 years. McCormick, using macroeconomic and quantitative analyses, detected growing stresses in the global economy before the meltdown." The reader, by looking at this, may understand that on one side ``mathematical models of quantitative analysis" (a sentence that sounds quite redundant) fail in times of crises, whereas ``macroeconomic and quantitative analyses" helped predicting some aspects of the crisis. One is left to wonder what is the different use of ``quantitative" between ``mathematical models of quantitative analysis" and ``macroeconomic and quantitative analyses". It is as though mathematics had all of a sudden become a bad word. Of course the article aims at saying that macroeconomic analysis and fundamentals are of increasing importance and should be taken more into account, although in our opinion it does not distinguish clearly valuation from prediction, but some of the sentences used to highlight this idea are quite symptomatic of the attitude we described above towards modeling and mathematics.

Another article that brings mathematics and mathematicians (provided that is what one means by ``math wizards") into the picture for the blaming is Lohr (2009), in ``Wall Street's Math Wizards Forgot a Few Variables", appeared in the New York Times of September 12. Also, Turner\footnote{Turner, J.A. (2009). The Turner Review. March 2009. Financial Services Authority, UK.\\ \url{www.fsa.gov.uk/pubs/other/turner\_review.pdf}.} (2009) has a section entitled ``Misplaced reliance on sophisticated maths".

This overall hostility and blaming attitude towards mathematics and mathematicians, whether in the industry or in academia, is the reason why we feel it is important to point out the following: the notion that even more mathematically oriented quants have not been aware of the Gaussian Copula\index{Gaussian Copula} model limitations is simply false, as we are going to show, and you may quote us on this. The quant and academic communities have produced and witnessed a large body of research questioning the copula assumption. This is well documented: there is even a book\footnote{Lipton, A. and Rennie, A. (Editors), Credit Correlation - Life After Copulas, World Scientific, 2007.} based on a one-day conference hosted by Merrill Lynch in London in 2006, well before the crisis, and called ``Credit Correlation: Life after Copulas". This conference had been organized by practitioners. The ``Life after Copulas" book contains several attempts to go beyond the Gaussian Copula and implied correlation, most of which come from practitioners (and a few by academics). But that book is only a tip of the iceberg. There are several publications that appeared pre-crisis and that questioned the Gaussian Copula and implied correlation. For example, we warned against the dangers implicit in the use of implied correlation in our report ``Implied Correlation: A paradigm to be handled with care", that we posted in SSRN in 2006, again well before the crisis.

Still, it seems that this is little appreciated by some market participants, commentators, journalists, critics, politicians, and academics. There are still a number of people out there who think that a formula killed Wall Street.

This paper brings a little clarity by telling a true story of pre-crisis warnings and also of pre-crisis attempts to remedy the drawbacks of implied correlation. We do not document the whole body of research that has addressed the limits of base correlation and of the Gaussian Copula, but rather take a particular path inside this body, based on our past research, that we also update to see what our models tell us in-crisis.

To put our paper in a nutshell, we can say that it starts from the payoffs of CDOs, explaining how to write them and how they work. We then move to the introduction of the inconsistent Gaussian Copula model and the related implied correlations, both compound and base, moving then to the GPL model: an arbitrage-free dynamic loss model capable of consistently calibrating all the tranches across attachments and detachments for all the maturities at the same time. En passant, we also illustrate the Implied Copula, a method that can consistently account for CDOs with different attachment and detachment points but not for different maturities, and the Expected Tranche Loss (ETL)\index{ETL} surface, a model independent approach to CDO prices interpolation.

We will see that, already pre-crisis, both the Implied Copula and the dynamic loss model imply modes far down the right tail of the loss distribution. This means that there are default probability clusters corresponding to joint default of a large number of entities (sectors) of the economy.\index{Cluster Defaults!sectors}

The discussion is abundantly supported by market examples through history. We cannot stress enough that the dangers and critics we present to the use of the Gaussian Copula and of implied correlation, and the modes in the tail of the loss distribution obtained with consistent models, had all been published by us, among others, in 2006, well before the crisis.

Despite these warnings, the Gaussian Copula model is still used in its base correlation formulation, although under some possible extensions such as random recovery.\index{Recovery!random} The reasons for this are complex. First the difficulty of all the loss models, improving the consistency issues, in accounting for single name data and to allow for single name sensitivities. This is due to the fact that if we model the loss of the pool directly as an aggregate object, without taking into account single defaults, then the model sensitivities to single name credit information are not in the picture. In other terms, while the aggregate loss is modeled so as to calibrate satisfactorily indices and tranches, the model does not see the single name defaults but just the loss dynamics as an aggregate object. Therefore partial hedges with respect to single names are not possible.
As these issues are crucial in many situations, the market practice remains with base correlation. Furthermore, even the few models achieving single name consistency have not been developed and tested enough to become operational on a trading floor or in a large risk management platform. Indeed, a fully operational model with realistic run times and numerical stability across a large range of possible market inputs would be more than a prototype with some satisfactory properties that has been run in some ``off-line" studies. Also, when one model has been coded in the libraries of a financial institution, changing the model implies a long path involving a number of issues that have little to do with modeling and more to do with IT problems, integration with other systems, and the likes. Therefore, unless a new model appears to be really promising and extremely convincing in all its aspects, there is reluctance in adopting it on the trading floor or on risk management systems.

Overall we conclude that the modeling effort in this area of the derivatives market is unfinished, partly for the lack of an operationally attractive single-name consistent dynamic loss model, and partly because of the diminished investment in this research area, but the fact that the modeling effort is unfinished does not mean that the quant community has been unaware of model limitations, as we abundantly document, and, although our narrative ends with an open finale, we still think it is an entertaining true story.

\subsection*{Part 2: How I learned to stop worrying and love the CDOs}
\addcontentsline{toc}{subsection}{Part 2: How I learned to stop worrying and love the CDOs}

We cannot close this preface without going back to the large picture, and ask the more general question: is the crisis due to poor modeling?

As we have seen, the market has been using simplistic approaches for credit derivatives, but it has also been trying to move beyond those. However, we should also mention that CDOs are divided into two categories: Cash\index{CDO!cash} and Synthetics\index{CDO!synthetic}. Cash CDOs involve hundreds or even thousands of names and have complex path-dependent payouts (``waterfalls"). Even so, Cash CDOs are typically valued by resorting to single homogeneous default-rate scenarios or very primitive assumptions, and very little research and literature is available on them. Hence these are complex products with sophisticated and path-dependent payouts that are often valued with extremely simplistic models. Synthetic CDOs are the ones we described in this preface and that will be addressed in this paper. They have more simple and standardized payouts than the cash CDOs but are typically valued with more sophisticated models, given the larger standardization and the ease in finding market quotes for their prices. Synthetic CDOs on corporates are epitomized by the quoted tranches of the standard pools DJ-iTraxx (Europe) and CDX (USA). However, CDOs, especially Cash, are available on other asset classes, such as loans (CLO)\index{CLO}\index{Collateralized Loan Obligation|see{CLO}}, residential mortgage portfolios (RMBS)\index{RMBS}\index{Residential Mortage-Backed Security|see{RMBS}}, commercial mortgages portfolios (CMBS)\index{CMBS}\index{Commercial Mortage-Backed Security|see{CMBS}}, and on and on. For many of these CDOs, and especially RMBS, quite related to the asset class that triggered the crisis, the problem is in the data rather than in the models. Bespoke corporate pools have no data from which to infer default ``correlation" and dubious mapping methods are used. At times data for valuation in mortgages CDOs (RMBS and CDO of RMBS)
are dubious and can be distorted by fraud\footnote{See for example the FBI Mortgage fraud report, 2007,\\ \url{www.fbi.gov/publications/fraud/mortgage$\_$fraud07.htm}.}.

At times it is not even clear what is in the portfolio: the authors have visioned offering circulars of a RMBS on a huge portfolio of residential mortgages where more than the $2\%$ of properties in the portfolio were declared to be of unknown type. What inputs can we give to the models if we do not even know the kind of residential property that functions as underlying of the derivative?

All this is before modeling. Models obey a simple rule that is popularly summarized by the acronym GIGO (Garbage In $\rightarrow$ Garbage Out). As Charles Babbage (1791-–1871) famously put it:

\begin{quote}
{\emph{
On two occasions I have been asked [by members of Parliament],
``Pray, Mr. Babbage, if you put into the machine wrong figures, will the right answers come out?"
I am not able rightly to apprehend the kind of confusion of ideas that could provoke such a question.
}}
\end{quote}

So, in the end, is the crisis due to models inadequacy? Is the crisis due to quantitative analysts and academics pride and unawareness of models
limitations?

We show in this paper that quants have been aware of the limitations and of extreme risks before the crisis. Lack of data or fraud-corrupted data, the fragility in the ``originate to distribute" system, liquidity and reserves policies, regulators lack of uniformity, excessive leverage and concentration in real estate investment, poor liquidity risk management techniques, accounting rules and excessive reliance on credit rating  agencies are often factors not to be underestimated. This crisis is a quite complex event that defies witch-hunts, folklore and superstition.
Methodology certainly needs to be improved but blaming just the models for the crisis appears, in our opinion, to be the result of a very limited point of view.

\bigskip

{\emph{London, Milan, Madrid, Pavia and Venice, December 1, 2009.}}

\medskip

Damiano Brigo, Andrea Pallavicini and Roberto Torresetti.

\bigskip

\subsection*{Acknowledgments}
We are grateful to Tom Bielecki for several interesting discussions and helpful correspondence. We are also grateful to Andrea Prampolini for helpful interaction on recovery modeling and other issues on the credit
derivatives market. Frederic Vrins corresponded with us on the early versions, helping with a number of issues.

Damiano wishes to express gratitude to co-authors, colleagues and friends who contributed to his insight in the last years, including Aurelien Alfonsi, Agostino Capponi, Naoufel El-Bachir, Massimo Morini, and the Fitch Solutions London team including Imane Bakkar, Johan Beumee,  Kyriakos Chourdakis, Antonio Dalessandro, Madeleine Golding, Vasileios Papatheodorou,  Ed Parcell, Mirela Predescu, Karl Rodolfo, Daniel Schiemert, Gareth Stoyle,  Rutang Thanawalla, Fares Triki, Weike Wu. This paper (and the related book) preparation has also been intersecting Damiano's wedding, so that he is especially grateful to Valeria for her patience, the wonderful wedding and the Patagonia honeymoon, and to both families for constant support and affection in difficult times.

Andrea is grateful to his colleagues and friends for their helpful contributions and patience.

Roberto thanks Luis Manuel Garc\'{\i}a Mu\~{n}oz and Seivane Navia Soledad for helpful suggestions and support with analysis and insight.

\newpage

\section[Introduction: credit modeling pre- and in-crisis]{Introduction:\\ credit modeling pre- and in-crisis}

This paper aims at showing the limits of popular models or pseudo-models (mostly quoting mechanisms with modeling semblance) that in the past years have been extensively used to mark to market and risk manage multi-name credit derivatives. We present a compendium of results we first published before the crisis, back in 2006, pointing out the dangers in the modeling paradigms used at the time in the market, and showing how the situation has even worsened subsequently by analyzing more recent data. We also point out that the current paradigm had been heavily criticized before the crisis, referring to our and other authors works addressing the main limitations of the current market paradigm well before popular accounts such as Salmon (2009) appeared.

Problems of the current paradigm include

\begin{itemize}
\item Unrealistic Gaussian copula assumption and flattening of 7750 pairwise dependence parameters into one.
\item Lack of consistency of the implied correlation market models with more than one tranche quote at the time
\item Occasional impossibility of calibration even of single tranches, or possibility to obtain negative expected tranched losses violating the arbitrage free constraints.
\item Lack of an implied loss distribution consistent with market CDO tranche quotes for a single maturity.
\item Lack of a loss distribution dynamics consistent with CDO tranche quotes on several maturities;
\end{itemize}

In this respect we will introduce examples of models published before the crisis that partly remedy the above deficiencies. All the discussion is supported by examples based on market data, pre- and in- crisis. In addressing these issues we adopt the following path through the different methodologies.

\subsection*{Bottom-up models}
A common way to introduce dependence in credit derivatives modeling is by means of copula functions. A typically Gaussian copula is postulated on the exponential random variables triggering defaults of the pool names according to first jumps of Poisson processes. In general, if one tries to model dependence by specifying dependence across single default times, one is in the so called ``bottom-up" framework, and the copula approach is typically within this framework.  Such procedure cannot be extended in a simple way to a fully dynamical model in general. We cannot do justice to the huge copula literature in credit derivatives here. We only mention that there have been attempts to go beyond the Gaussian copula introduced in the CDO world by
Li (2000) and leading to the implied (base and compound) correlation framework, some important limits of which have been pointed out in Torresetti et al. (2006b). Li et al (2005) also proposed a mixture approach in connection with CDO squared. For results on sensitivities computed with the Gaussian copula models see for example Meng and Sengupta (2008).

An alternative to copulas in the bottom up context is to insert dependence among the default intensities of single names, see for example the paper by Chapovsky, Rennie and Tavares (2006). Joshi and Stacey (2006) resort to modeling business time to create default correlation in otherwise independent single names defaults, resorting to an ``intensity gamma" framework. Similarly but in a firm value inspired context, Baxter (2006) introduces Levy firm value processes in a bottom up framework for CDO calibration. Lopatin (2008) introduces a bottom up framework effective in the CDO context as well, having single name default intensities being deterministic functions of time and of the pool default counting process, then focusing on hedge ratios and analyzing the framework from a numerical performances point of view, showing this model to be interesting even if lacking explicit modeling of single names credit spread volatilities\index{Credit Spread!volatility}.


Going back to bottom-up models in the context of CDOs, Albanese et al. (2006) introduce a bottom-up approach based on structural models ideas that can be made consistent with several inputs both under the historical and pricing measures and that manages to calibrate CDO tranches.


\subsection*{Compound correlation}

Building on Torresetti et al (2006b), in the context of bottom-up models, we start with the net present value (NPV) of synthetic Collateralized Debt Obligations (CDO) tranches on pools of corporate credit references in its original layout: the compound correlation framework.

We highlight two of the major weaknesses of the compound correlation:
\begin{itemize}
\item Lack of robustness of the compound correlation framework in view of the non-invertibility of mainly the 10-year-maturity DJi-Traxx $6-9\%$ and CDX $7-10\%$ tranches and more recently the non-invertibility of mainly the 10-year-maturity DJi-Traxx $12-22\%$ and CDX $10-15\%$ tranches.
\item Flattening information on 7750 pairwise correlation parameters into a single one for each tranche.
\item More importantly from a practical standpoint, we highlight the typical non smooth behaviour of the compound correlation and the resulting difficulties in pricing bespoke CDO tranches.
\end{itemize}

\subsection*{Base correlation}

We then introduce the next step the industry took, see for example McGinty and Ahluwalia (2004), namely the introduction of base correlation, as a solution to both problems given the fact that:
\begin{itemize}
\item	the resulting map is much smoother, thus facilitating the pricing of bespoke tranche spreads from liquid index tranches;
\item	until early 2008 the heterogeneous pool one-factor Gaussian copula base correlation has been consistently invertible from index market tranche spreads
\end{itemize}

Nevertheless we expose what are some of the known remaining weaknesses of the base correlation framework:

\begin{itemize}
\item	Depending on the interpolation technique being used, tranche spreads could be not arbitrage free. In fact for senior tranches it may well be that the expected tranche loss plotted versus time is initially decreasing.
\item	The impossibility of inverting correlation for senior AAA and super senior tranches.  
\item Inconsistency at single tranche valuation level, as two components of the same trade are valued with models having two different parameter values;
\item Last but not least, flattening information on 7750 pairwise correlation parameters into a single one for each equity tranche trade.
\end{itemize}

As an explanation to the first weakness we point to the fact related to the third one, namely that this arises because the NPV of each tranche is obtained computing the expected tranche loss and outstanding notional under two different distributions (the distribution corresponding to the attachment base correlation and the one corresponding to the detachment base correlation) so that base correlation is an inconsistent notion already at single tranche level.

As an explanation to the second weakness we point to the fact that the deterministic recovery assumption, whilst being computationally very convenient, does not allow to capture the more recent market conditions. This has been addressed in the implied correlation framework by Amraoui and Hitier (2008) and Krekel (2008). However, even with this update, base correlation remains exposed to the remaining three weaknesses. 


%
%
%

Base correlation, with updates and variants, remains to this day the main pricing method for synthetic corporate CDOs, regardless of the body of research criticizing it we hint at above and below.

\subsection*{Implied copula}

We next summarize the concept of Implied Copula (introduced by Hull and White (2006) as ``Perfect Copula") as a non-parametric model, to deduce from a set of market CDO spreads, spanning the entire capital structure,  the shape of the risk-neutral pool loss distribution. The general use of flexible systemic factors has been later generalized and vastly improved by Rosen and Saunders (2009), who also discuss the dynamic implications of the systemic factor framework. Factors and dynamics are also discussed in Inglis et al. (2008), while Eberlein, Frey and von Hammerstein (2008) generalize the original factor model by Vasicek (1987, 1991) and also propose a dynamic Markov chain model for CDO tranche pricing.

Our calibration results based on the implied copula, already seen in Torresetti et al (2006c), point out that a consistent loss distribution across tranches for a single maturity features modes in the tail of the loss distribution. These probability masses on the far right tail imply default possibilities for large clusters (possibly sectors) of names of the economy. These results had been published originally in 2006 on SSRN.com. We will report such features here and we will find the same features again following a completely different approach below.

Here we highlight the persistence of the modes (bumps) in the right tail of the implied loss distribution
\begin{itemize}
\item	Through time, via historical calibrations
\item	Through regions, comparing the results of the historical calibration to the DJi-Traxx and the CDX
\item	Through maturities, comparing the results of the calibration to different maturities
\end{itemize}

The Implied Copula can calibrate consistently across the capital structure\index{Implied Copula!consistency} but not across maturities, as it is a model that is inherently static. The next step thus consists in introducing a dynamic loss model. This moves us into the so called top-down framework (although dynamic approaches are also possible in the bottom-up context, as we have seen in part of the above references). But before analyzing the top-down framework in detail, we make a quick diversion for a model-independent approach to CDO tranches pricing and interpolation.


\subsection*{Expected Tranche Loss (ETL) Surface}

Expected tranche losses (ETL)\index{ETL}\index{Expected Tranche Loss|see{ETL}} for different detachment points and maturities can be viewed as the basic bricks on which synthetic CDO formulas components are built with linear operations (but under some non-linear constraints). We explain in detail how the payoffs of credit indices and tranches are valued in terms of expected tranched losses (ETL). This methodology, first illustrated pre-crisis in Torresetti et al. (2006a), reminds of Walker's (2006) earlier work and of the formal analysis of the properties of expected tranche loss in connection with no arbitrage in Livesey and Schl{\"o}gl (2006).

ETL are natural quantities to imply from market data. No-arbitrage constraints on ETL's as attachment points and maturities change are introduced briefly. As an alternative to the inconsistent notion of implied correlation illustrated earlier, we consider the ETL surface, built directly from market quotes given minimal interpolation assumptions\index{ETL!consistency}. We check that the kind of interpolation does not interfere excessively with the results. Instruments bid/asks enter our analysis, contrary to Walker's (2006) earlier work on the ETL implied surface. By doing so we find less violations of the no-arbitrage conditions.

We also mention some further references appeared later and dealing with evolutions of this technique: Parcell and Wood (2007), again pre-crisis, consider carefully the impact of different kinds of interpolation, whereas Garcia and Goossens (2007) compare ETL between the Gaussian copula and L\'evy models.

In general the ETL implied surface can be used to value tranches with nonstandard attachments and maturities as an alternative to implied correlation. However, deriving hedge ratios as well as extrapolation may prove difficult. Also, ETL is not really a model but rather a model-independent stripping algorithm, although the particular choice of interpolation may be viewed as a modeling choice. Eventually ETL is not helpful for pricing more advanced derivatives such as tranche options or cancelable tranches. This is because ETL does not specify an explicit dynamics for the loss of the pool. To that we turn now, by looking at the top-down dynamic loss models.

\subsection*{Top (down) framework}

One could give up completely single name default modeling and focus on the pool loss and default counting processes, thus considering a dynamical model at the aggregate loss level, associated to the loss itself or to some suitably defined loss rates. This is the ``top-down" approach, see for example Bennani (2005, 2006), Giesecke, Goldberg and Ding (2005), 
Sch\"onbucher (2005), Di Graziano and Rogers (2005), Brigo, Pallavicini and Torresetti (2006a,b), Errais, Giesecke and Goldberg (2006), Lopatin and Misirpashaev (2007), Ding, Giesecke and Tomecek (2009) among others.

The first joint calibration results of a dynamic loss model across indices, tranches attachments and maturities, available in Brigo, Pallavicini and Torresetti (2006a), show that even a relatively simple loss dynamics, like a capped generalized Poisson process, suffices to account for the loss distribution dynamical features embedded in market quotes.

This work also confirms the implied-copula findings of Torresetti et al (2006c), showing that the loss distribution tail features a structured multi-modal behaviour implying non negligible default probabilities for large fractions of the pool of credit references, showing the potential for high losses implied by CDO quotes before the beginning of the crisis. Cont and Minca (2008) use a non-parametric algorithm for the calibration of top models, constructing a risk neutral default intensity process for the portfolio underlying the CDO, looking for the risk neutral loss process ``closest" to a prior loss process using relative entropy techniques. See also Cont and Savescu (2008).

However, in general to justify the ``down" in ``top-down" one needs to show that from the aggregate loss model one can recover a posteriori consistency with single-name default processes when they are not modeled explicitly. Errais, Giesecke and Goldberg (2006) advocate the use of random thinning techniques for their approach, see also Halperin and Tomecek (2008), who delve into more practical issues related to random thinning of general loss models, and Giesecke, Goldberg and Ding (2005) who compare the thinning based edges of the top down model with the copula-based ones. Bielecki, Crepey and Jeanblanc (2008) build semi-static hedging examples and consider cases where the portfolio loss process may not be a sufficient statistics.

Still, it is not often clear for specific models whether a fully consistent single-name default formulation is possible given an aggregate model as the starting point. There is a special ``bottom-up" approach that can lead to a distinct and rich loss dynamics. This approach is based on the common Poisson shock (CPS) framework, reviewed in Lindskog et al. (2003). This approach allows for more than one defaulting name in small time intervals, contrary to some of the above-mentioned ``top-down" approaches. In the ``bottom-up" language, one sees that this approach leads to a Marshall-Olkin copula linking the first jump (default) times of single names.  In the ``top-down" language, this model looks very similar to the Generalized Poisson Loss model in Brigo et al (2006a) when one does not cap the number of defaults. The problem of the CPS framework is that it allows for repeated defaults,
which is clearly wrong as one name could default more than once.

In the credit derivatives literature the CPS framework has been used for example in Elouerkhaoui (2006), see also references therein. Balakrishna (2006) introduces a semi-analytical approach allowing again for more than one default in small time intervals and hints at its relationship with the CPS framework, showing also some interesting calibration results. Balakrishna (2007) then generalizes this earlier paper to include delayed default dependence and contagion.

\subsection*{Generalized Poisson (Cluster) Loss model}

Brigo et al (2007) address the repeated default issue in CPS by controlling the clusters default dynamics to avoid repetitions. They calibrate the obtained model satisfactorily to CDO quotes across attachments and maturities, but  the combinatorics for a non-homogeneous version of the model are forbidding, and the resulting GPCL approach is hard to use successfully in practice when taking into account single names. Still, in the context of the present paper, the GPL and GPCL models will be useful in showing how a loss distribution dynamics consistent with CDO market quotes should evolve.

In this paper we summarize the Generalized Poisson Loss model, leaving aside the GPCL model. As explained above, GPL is a dynamical models for the loss, able to reprice all tranches and all maturities at the same time. We employ here a variant that models directly the loss rather than the default counting process plus recovery. The loss is modeled as the sum of independent Poisson processes, each associated with the default of a different number of entities, and capped at the pool size to avoid infinite defaults.  The intuition of these driving Poisson processes is that of defaults of sectors, although the sectors amplitudes vary in our formulation of the model pre- and in-crisis. In the new model implementation in-crisis for this paper we fix the amplitude of the loss triggered by each cluster of defaults a priori, without calibrating it as we were doing in our earlier GPL work. This makes the calibration more transparent and the calibrated intensities of the sectors defaults easier to interpret. We point out, however, that the precise default of sectors is made rigorous only in GPCL.

We highlight how the GPL model is able to reproduce the tail multimodal feature that the Implied Copula proved to be indispensable to reprice accurately the market spreads of CDO tranches on a single maturity. We also refer to the later related results of Longstaff and Rajan (2007), that point in the same direction but adding a principal component analysis on a panel of CDS spread changes, with some more comments on the economic interpretation of the default clusters being sectors. An econometric investigation of cluster defaults starting from the Poisson framework is in Duan (2009).

\begin{remark}
We draw the reader's attention to the default history, pointing to default clusters being concentrated in a relatively short time period (a few months) like the thrifts in the early 90s at the height of the loan and deposit crisis, airliners after 2001, autos and financials more recently. In particular, from the 7th September 2008 to the 8th October 2008, a time window of one month, we witnessed seven credit events occurring to major financial entities: Fannie Mae, Freddie Mac, Lehman Brothers, Washington Mutual, Landsbanki, Glitnir, Kaupthing.  Fannie Mae and Freedie Mac conservatorships were announced on the same date (September 7, 2008) and the appointment of a ``receivership committee" for the three icelandic banks (Landsbanki, Glitnir, Kauping) was announced between the 7th and the 8th of October.
\end{remark}

\begin{remark}
Standard and Poors issued a request for comments related to changes in the rating criteria of corporate CDO\footnote{see ``Request for Comment: Update to Global Methodologies and Assumptions for Corporate Cash Flow CDO and Synthetic CDO Ratings" , 18-Mar-09, Standard \& Poor's.}.  Thus far agencies have been adopting a multifactor Gaussian Copula approach to simulate the portfolio loss in the objective measure.  S\&P proposed changing the criteria so that tranches rated 'AAA' should be able to withstand the default of the largest single industry in the asset pool with zero recoveries.  We believe this goes in the direction of modelling the loss in the risk neutral measure via GPL like processes, given that this implies admitting as a stressed but plausible scenario the possibility that a  cluster defaults in the objective measure. See also Torresetti and Pallavicini (2007) for the specific case of Constant Proportion Debt Obligations (CPDO).
\end{remark}


We finally comment more generally on the dynamical aggregate models and on their difficulties to lead to single name hedge ratios when trying to avoid complex combinatorics. The framework remains thus incomplete to this day, because obtaining jointly tractable dynamics and consistent single name hedges, that can be realistically applied in a trading floor, remains a problem. We provided some references for the latest research in this field above. We highlight, though, that even a simple dynamical model like our GPL or the single-maturity implied copula is enough to appreciate that the market quotes were implying the presence of large default clusters with non-negligible probabilities well in advance of the credit crisis, as we documented in 2006 and early 2007.


\subsection*{Paper structure}


Section \ref{sec:market} introduces the index and CDO tranche payouts we will be analyzing in this paper, explaining also the definition of tranche spread and upfront quotes.
Section~\ref{sec:gausscop} introduces the Gaussian copula model, in its different formulations concerning homogeneity and finiteness, and then illustrates the notions of implied correlation from CDO tranche quotes. The two paradigms of base correlation and compound correlation are explained in detail. Existence and uniqueness of implied correlation are discussed on a number of market examples, highlighting the pros and cons of compound and base correlations, and the limitations inherent in these concepts. The section ends with a summary of issues with implied correlations, pointing out the danger for arbitrage when negative expected tranche losses surface, and the lack of consistency across capital structure and maturity. The first inconsistency is then addressed in Section~\ref{sec:impliedcopula}, with the implied copula, illustrated with a number of studies throughout a long period, both pre- and in- crisis, whereas both inconsistencies are addressed in Section~\ref{sec:gplmodel}, where our full-fledged GPL dynamic loss model is illustrated pre-crisis. In Section~\ref{sec:gplmodel} we explore en passant a model-free extraction of expected loss information from CDO quotes that can be occasionally helpful in interpolating or checking arbitrage constraints. All these paradigms are then analyzed in crisis in Section~\ref{sec:recent}, while the final discussion, including the reasons why implied correlation is still used despite all its important shortcomings, are given in Section~\ref{sec:conclusions}. In Particular, the need for hedge ratios with respect to single names, random recovery modeling and speed of calibration remain issues that are hard to address jointly outside the base correlation framework.


\section{Market quotes\label{sec:market}}

For single names our reference products will be credit default swaps (CDS).

The most liquid multi-name credit instruments available in the market are instead credit indices and CDO tranches (e.g. DJi-TRAXX, CDX). We discuss them in the following.

The procedure for selecting the standardized pool of names is the same for the two indices. Every six months a new series is rolled at the end of a polling process managed by MarkIt where a selected list of dealers contributes the ranking of the most liquid CDS. All credit references that are not investment grade are discarded.  Each surviving credit reference underlying the CDS is assigned to a sector.  Each sector is contributing a predetermined number of credit references to the final pool of names. The rankings of the various dealers for the investment grade names are put together to rank the most liquid credit references within each sector.

The index is given by a pool of names $1,2,\ldots,M$, typically $M=125$, each with notional $1/M$ so that the total pool has unitary notional. The index default leg consists of protection payments corresponding to the defaulted names of the pool. Each time one or more names default the corresponding loss increment is paid to the protection buyer, until final maturity $T=T_b$ arrives or until all the names in the pool have defaulted.

In exchange for loss increase payments, a periodic premium with rate $\cds$ is paid from the protection buyer to the protection seller, until final maturity $T_b$. This premium is computed on a notional that decreases each time a name in the pool defaults, and decreases of an amount corresponding to the notional of that name (without taking out the recovery).

We denote with $\lossnorm_t$ the portfolio cumulated loss and with $\defratenorm_t$ the number of defaulted names up to time $t$ divided by $M$. Since at each default part of the defaulted notional is recovered, we have $0 \le d\lossnorm_t \le d\defratenorm_t \le 1$. The discounted payoff of the two legs of the index is given as follows:
\[
\defleg(0) := \int_0^T D(0,t) d\lossnorm_t
\]
\[
\prmleg(0) = \cds_0 \sum_{i=1}^b \,\delta_i D(0,T_i) (1-\defratenorm_{T_i})
%
\]
where $D(s,t)$ is the discount factor (often assumed to be deterministic) between times $s$ and $t$ and $\delta_i=T_i-T_{i-1}$ is the year fraction. In the second equation the actual outstanding notional in each period would be an average over $[T_{i-1},T_i]$, but we replaced it with the value of the outstanding notional at $T_i$ for simplicity.

The market quotes the values of $\cds_0$ that, for different maturities, balances the two legs. Assuming deterministic default-free interest rates, if one has a model for the loss and the number of defaults one may impose that the loss and number of defaults in the model, when plugged inside
the two legs, lead to the same risk neutral expectation (and thus price)

\begin{eqnarray}
\label{eq:index}
\cds_0 = \frac{ \int_0^T D(0,t) \,d \Ex{0}{ \lossnorm_t } }
        { \sum_{i=1}^b \,\delta_i D(0,T_i) ( 1 - \Ex{0}{ \defratenorm_{T_i} } ) }
\end{eqnarray}

Synthetic CDO with maturity $T$ are contracts involving a protection buyer, a protection seller and an underlying pool of names. They are obtained by  ``tranching" the loss of the pool between the points $A$ and $B$, with $0 \le A<B \le 1$.
\[
\losstr_t:= \frac{1}{B-A}\left[(\lossnorm_t-A) \ind{A<\lossnorm_t\le B}+(B-A)\ind{\lossnorm_t>B}\right]
\]
An alternative expression that is useful is
\begin{equation}\label{lossasdifferenceeq}
\losstr_t:= \frac{1}{B-A}\left[B \lossnorm^{0,B}_t -A \lossnorm^{0,A}_t\right]
\end{equation}

Once enough names have defaulted and the loss has reached $A$, the count starts. Each time the loss increases the corresponding loss change re-scaled by the tranche thickness $B-A$  is paid to the protection buyer, until maturity arrives or until the total pool loss exceeds $B$, in which case the payments stop.

The discounted default leg payoff can then be written as
\[
\defleg_{A,B}(0) := \int_0^T D(0,t) d\losstr_t
\]
Again, one should not  be confused by the integral, the loss $\losstr_t$ changes with discrete jumps. Analogously, also the total loss $\lossnorm_t$ and the tranche outstanding notional change with discrete jumps.

As usual, in exchange for the protection payments, a premium rate $\cdstr_0$, fixed at time $T_0=0$, is paid periodically, say at times $T_1,T_2,\ldots,T_b=T$. Part of the premium can be paid at time $T_0=0$ as an upfront $\upfronttr_0$. The rate is paid on the ``survived" average tranche notional. If we further assume payments are made on the notional remaining at each payment date $T_i$, rather than on the average in $[T_{i-1},T_i]$, the premium leg can be written as
\begin{align*}
\prmleg_{A,B}(0) & := \upfronttr_0 +  \cdstr_0  \dv_{A,B}(0)  \\
\dv_{A,B}(0) & :=  \sum_{i=1}^b \, \delta_i D(0,T_i) (1-\losstr_{T_i})
\end{align*}

When pricing CDO tranches, one is interested in the premium rate $\cdstr_0$ that sets to zero the risk neutral price of the tranche. The tranche value is computed taking the (risk-neutral) expectation (in $t=0$) of the discounted payoff consisting on the difference between the default and premium legs above. Assuming deterministic default-free interest rates we obtain
\begin{eqnarray}
\label{eq:tranche}
\cdstr_0 = \frac{ \int_0^T D(0,t) d \Ex{0}{ \losstr_t } - \upfronttr_0 }
                { \sum_{i=1}^b \, \delta_i D(0,T_i) (1 - \Ex{0}{ \losstr_ {T_i} } ) }
\end{eqnarray}
%
The above expression can be easily recast in terms of the upfront premium $\upfronttr_0$ for tranches that are quoted in terms of upfront fees.

The tranches that are quoted on the market refer to standardized pools, standardized attachment-detachment points $A-B$ and standardized maturities.  The standardized attachment and detachment are slightly different for the CDX.NA.IG and the DJi-Traxx Europe Main\footnote{The attachment of the CDX tranches are slightly higher reflecting the average higher perceived riskiness (as measured for example by the CDS spread or balancesheet ratios) of the liquid investment grade north american names.}.

\begin{table}[t]
\centering
\begin{tabular}{cc}
\hline \hline
DJi-Traxx Europe Main &  CDX.NA.IG  \\
\hline
    0-3\%     &    0-3\%       \\
    3-6\%     &    3-7\%       \\
    6-9\%     &    7-10\%       \\
    9-12\%    &    10-15\%       \\
    12-22\%   &    15-30\%       \\
    22-100\%  &    30-100\%       \\
\hline
\end{tabular}
\caption{\label{tab:stdAttDet} Standardized attachment and detachment for the DJi-Traxx Europe Main and CDX.NA.IG tranches}
\end{table}

For the DJi-Traxx and CDX pools, the equity tranche $(A=0, B=3\%)$ is quoted by means of the fair $\upfronttr_0$, while assuming\footnote{The reason for the equity tranche to be quoting as upfront is to reduce the counterparty credit risk the protection seller is facing.} $\cdstr_0 = 500 bps$.  All other tranches are generally quoted by means of the fair running spread $\cdstr_0$, assuming no upfront fee  ($\upfronttr_0=0)$.   Following the recent market turmoil also the 3-6\% and the 3-7\% have been quoting in terms of an upfront amount and a running $\cdstr_0 = 500 bps$ given the exceptional riskiness priced by the market also for mezzanine tranches.

\section{Gaussian copula model}\label{sec:gausscop}

The Gaussian Copula model is a possible way to model dependence of random variables and, in our case, of default times.  As the default event of a credit reference is a random binary variable, the correlation between default events is not an intuitive object to handle. We need to focus our attention rather on default times. We denote by $\tau_i$ the default time of name $i$ in a pool of $M$ names. Default times of different names need to be connected. The copula formalism allows to do this in the most general way.

Indeed, if $p_i(t) = \Qx{\tau_i \le t}$ is the default probability of name $i$ by time $t$, we know that the random variable $p_i(\tau_i) = U_i$ is a uniform random variable. Copulas are multivariate distributions on uniform random variables. If we call $C(u_1,\ldots,u_n)$ a multivariate uniform distribution, and $U_1,\ldots, U_M$ is a multivariate uniform with distribution $C$, then a possible multivariate distribution of the default times with marginals $p_i$ is
\[ \tau_1 := p_1^{-1}(U_1), \ldots, \tau_M := p_M^{-1}(U_M),\]
where for simplicity we are assuming the $p$s to be strictly invertible.
Clearly, since the $U_1,\ldots,U_M$ variables are connected through a multivariate distribution $C$, we have a dependence structure on the default times.

The Gaussian copula enters the picture when we assume that
\[ [U_1,\ldots,U_M] = [\Phi(X_1), \cdots, \Phi(X_M)] \]
where the $X_i$ are standard Gaussian random variables and $[X_1,\ldots,X_M]$ is a given multivariate Gaussian random variable with a given correlation matrix. $\Phi$ is the cumulative distribution function of the one-dimensional standard Gaussian. In the Gaussian Copula model the default times are therefore linked via normally distributed latent factors $X$.

A particular structure is assumed for the default probabilities $p_i$. The default probabilities of single names are supposed to be related to hazard rates $\lambda$. In other terms $p_i(t) = 1- \exp(-\int_0^t \lambda_i(s) ds)$. We define $\Lambda_i(t):=\int_0^t \lambda_i(s) ds$.

We also anticipate that the correlation matrix characterizing the Gaussian copula is often taken with all off-diagonal entries equal to each other according to a common value $\rho$ in corporate synthetic CDO valuation applications. This is always the case in particular when defining implied correlations.

$\cdstr_0$ (spread) or $\upfronttr_0$ (upfront) would be provided by the market and a correlation number $\rho$ characterizing a Gaussian copula with a correlation matrix where all entries are equal to $\rho$ (in this case we will say the correlation matrix to be ``flat to $\rho$") would be implied from the market quote. Indeed, Formula (\ref{eq:tranche}) defined the market quotes in terms of expectations of the tranced loss $\bar{L}^{A,B}$; in turn, the loss to be tranched at a given time is defined in terms of single default times as
\begin{equation}
\label{eq:poolLoss}
\lossnorm_t = \sum_{i=1}^M \frac{1}{M} (1-\rec_i) 1_{\{\tau_i \le t \}} = \sum_{i=1}^M \frac{1}{M} (1-\rec_i) 1_{\{\Phi(X_i) \le p_i(t) \}}
\end{equation}
where the $X_i$ are multivariate Gaussians with correlation matrix flat to $\rho$ and $\rec_i$ are the recovery rates associated to each name.

The correlation parameter $\rho$ is therefore affecting the tranche price since it is a key statistical parameter contributing to the tranched loss distribution whose expectation will be used in matching the market quote. Implied correlation aims at finding the value of $\rho$ consistent with a given tranche quote when every other model parameter (the $p_i$s, the $\rec_i$s) has been fixed.

We next illustrate the two main types of implied correlation one may obtain from market CDO tranche spreads:
\begin{enumerate}
\item Compound correlation is more consistent at single tranche level but for some market CDO tranche spreads cannot be implied or more than one correlation can be implied.
\item Base correlation is less consistent but more flexible and can be implied for a much wider set of CDO tranche market spreads. Furthermore, base correlation is more easily interpolated and leads to the possibility to price non-standard detachments.  Even so, Base correlation may lead to negative expected tranche losses, thus violating basic no-arbitrage conditions. We illustrate these features with numerical examples.
\end{enumerate}

We will first introduce the general One-Factor Gaussian Copula model.  Then we will introduce the Finite Pool (where $M$ is finite) Homogeneous ($p_1=p_2 =\ldots = p_M$ and $R_1=R_2=\ldots=R_M$) One-Factor Gaussian Copula model.  We will show how the loss probabilities formulas can be computed in this case.


\begin{remark}
In this work we stay with the homogeneous version, since this is enough to highlight some of the key flaws of implied correlation, leaving to the full book of Brigo, Pallavicini and Torresetti (2010) the detailed derivation of the formulas for the heterogeneous and large pool versions.  It is to be said that in practice, due to the necessity of computing hedge rations with respect to single names, the heterogeneous version is used, even if spread dispersion across names, leading possibly to very different $p_i$s, complicates matters.
\end{remark}

\subsection{One-factor Gaussian copula model}

A One-factor copula structure is a special case of the Gaussian copula above where
\[
X_i = \sqrt{\rho_i}\,S+\sqrt{1-\rho_i}\,Y_i,
\]
with $Y_i,S$ standard independent Gaussian variables.
$S$ is a systemic factor affecting default times of all names. $Y_i$ is a idiosyncratic factor affecting just the $i$-th name.

This parameterization would lead to a correlation between Gaussian factors $X_i$ and $X_j$ given by $\sqrt{\rho_i \rho_j}$. However, consistently with the flat correlation assumption and the homogeneity assumption all pairwise correlation parameters collapse to a single common value, i.e. $\rho_i = \rho$ for all $i$.

Assuming deterministic and possibly distinct recovery rate $R_i$ upon default we are able to simulate the pool loss at any time $t$ starting from the simulation of the Gaussian variables $X_i$s starting from equation \eqref{eq:poolLoss}.
Thus we are able to simulate the NPV of the premium and default leg of any tranche.


The heterogeneous Gaussian Copula model assumes possibly different recovery rates $\rec_i$ and probabilities of default $p_i(t)$.   The large pool assumption basically assumes homogeneity of both the recovery rates and probabilities of default and also assumes an infinite number of credit references $M$ leading to a substantial increase in the numerical efficienty of the pricing formulas.

In the following, as already mentioned, we stay with the homogeneous version thus assuming identical $\rec_i$ and $p_i(t)$ for all $M$ credit references in the pool.

Calculating the NPV of a derivative instrument via simulations can be necessary but may lead to intensive numerical effort.  Introducing the assumption of homogeneity it turns out that the Gaussian Copula model yields a semi-analytical formula to calculate the distribution of the pool loss.

Assume we know the realization of the systemic factor $S$.  In this case, conditional on $S$, the default events of the pool of credit references are independent.
The default by time $T$ conditional on the realization of the systemic factor $S$ of each single credit reference in the pool is a Bernoulli random variable with the same event probability of default
\begin{equation*}
\Qx{\tau_i < T |S} = \Phi \left( \frac{\Phi^{-1}( 1 - \exp(-\Lambda(T)))-\sqrt{\rho} S}{\sqrt{1-\rho}}\right).
\end{equation*}

The number of defaulted entities in the pool by time $T$ conditional on the realization of the systemic factor $S$ is the sum of $M$ Bernoulli variables and thus is binomially distributed.
\begin{equation}
\label{eq:binomPdm}
\Qx{\defratenorm_T = \frac{n}{M} \Big| S } = \frac{M!}{n!(M-n)!} \, \Qx{\tau_i < T |S}^n \left( 1 - \Qx{\tau_i < T |S} \right)^{M-n}
\end{equation}
where we point out that the $\Qx{\tau_i < T |S}$s are the same for all $i$s, since we are dealing with an homogeneous pool assumption.

We can now integrate the expression in (\ref{eq:binomPdm}) to get the unconditional probability of $n$ defaults occurring to the standardized pool of $M$ credit references before time $T$, leading to
\begin{equation}
\label{eq:integraM}
\Qx{\lossnorm_T=(1-\rec)\frac{n}{M}} = \int_{-\infty}^{+\infty} \Qx{\defratenorm_T = \frac{n}{M} \Big| S } \varphi(S) \, dS
\end{equation}
By computing the integral in (\ref{eq:integraM}) for $n=1,...,M$ we obtain the unconditional distribution of the pool loss rate that we need in order to compute the theoretical tranche spread $\cdstr_0$ (or upfront $\upfronttr_0$) in equation (\ref{eq:tranche}).

The integral in equation (\ref{eq:integraM}) does not allow for a closed form solution whereas all other quantities can be analytically calculated.  Hence the name semi-analytical (analytical up to the calculation of the integral) for the finite-pool homogeneous one-factor Gaussian copula model formula for theoretical tranche spread in (\ref{eq:tranche}).

\subsection{Compound correlation}\label{sec:compoundcorr}

Compound correlation is a first paradigm for implying credit default dependence from liquid market data. This approach consists in linking defaults across single names through a Gaussian copula where all the correlation parameters are collapsed to one.

Given this correlation parameter and given the desired declination of the One-Factor Gaussian Copula we can compute the loss distribution on a given set of dates.  Thus we can compute the expectations contained in (\ref{eq:tranche}), and with them the fair tranche spread.

A key step that will allow us to distinguish compound from base correlation is the decomposition of the tranche loss according to Equation (\ref{lossasdifferenceeq}). Since the default leg and premium leg (in particular the DV01) defining (\ref{eq:tranche}) are linear in the tranced loss $\losstr_t$,
by Equation (\ref{lossasdifferenceeq}) this results to be linear in base tranched losses $\lossnorm^{0,B}_t$ and $\lossnorm^{0,A}_t$.

Now one key step is that when we evaluate the expected $\losstr_t$ through (\ref{lossasdifferenceeq}), we can use two different copula correlations for the two pieces $\lossnorm^{0,A}_t$ (correlation $\rho_A$) and $\lossnorm^{0,B}_t$ (correlation $\rho_B$). As a consequence, the final formula reads

\begin{equation}\label{expectlossasdifferenceeq}
\mathbb{E}_0[\losstr_t](\rho_A,\rho_B):= \frac{1}{B-A}\left[B \mathbb{E}_0[\lossnorm^{0,B}_t](\rho_B) -A \mathbb{E}_0[\lossnorm^{0,A}_t(\rho_A)]\right]
\end{equation}
A similar decomposition holds for the DV01, that is a linear combination of such objects.

Consider the DJi-Traxx tranches for example, to clarify the procedure. For a given maturity in 3y, 5y, 7y, 10y, consider the market quotes
\[U^{0,3\%\ \mbox{\tiny Mkt} }+ 500bps  \mbox{\ running}, \]\[S^{3,6\%\ \mbox{\tiny Mkt}}, S^{6,9\%\
\mbox{\tiny Mkt}}, S^{9,12\%\ \mbox{\tiny Mkt}}, S^{12,22\%\
\mbox{\tiny Mkt}}\] To obtain the implied correlation one proceeds
as follows.

First solve in $\rho_{3\%}$ for the equity tranche
\[
\defleg_{0,3\%}(0, \rho_3, \rho_3) = 500bps  \dv_{0,3}(0, \rho_3, \rho_3) + U^{0,3\% \mbox{\tiny \ Mkt}}
\]
Then in moving on one has two choices: retain $\rho_3$ from the earlier tranche calibration and solve
\[
\defleg_{3,6}(0, \rho_3, \rho_6) = S^{3,6\%\mbox{\tiny \ Mkt}} \ \ \dv_{3,6}(0, \rho_3, \rho_6)
\]
in $\rho_6$ (base correlation) or solve
\[
\defleg_{3,6}(0, \bar{\rho}_{3,6}, \bar{\rho}_{3,6}) = S^{3,6\%\mbox{\tiny \ Mkt}} \ \ \dv_{3,6}(0,\bar{\rho}_{3,6}, \bar{\rho}_{3,6})
\]
in a new $\bar{\rho}_{3,6}$ (compound correlation). The next step will be again similar, and we iterate, until we reach the end of the capital structure.

\begin{table}[htp]
\centering
\begin{tabular}{ccc}
\hline \hline
  Tranche      &  Running  &   Upfront \\
\hline
0-3\%   &  500 bps  &  49\%   \\
3-6\%   &  360 bps  &  0\%    \\
6-9\%   &  82 bps   &  0\%    \\
9-12\%  &  46 bps   &  0\%    \\
12-22\% &  31 bps   &  0\%    \\
\hline
\end{tabular}
\caption{
\label{tab:tableCompInvert}DJi-Traxx Europe S5 10 year tranches quotes on August 3, 2005.
}
\end{table}

\begin{figure}[htp]
\begin{center}
\includegraphics[width=0.5\textwidth]{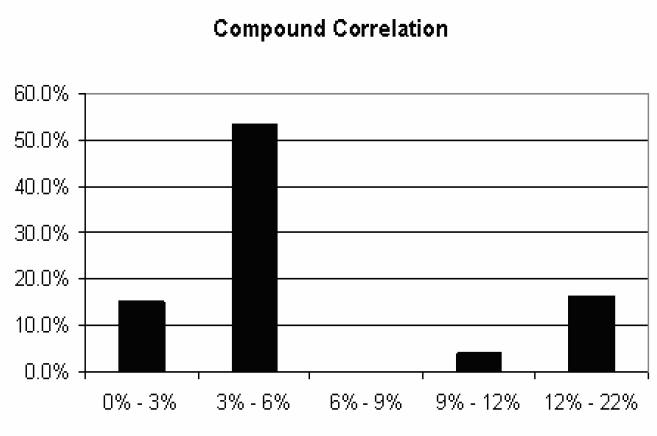}\vspace*{-1cm}
\end{center}
\caption{
\label{fig:comp_invert_skew}
Implied correlation smile calibrated to the DJi-Traxx S5 10 year tranche spreads on August 3, 2005.
}
\end{figure}

Compound correlation is more consistent at the level of single tranche, since we value the whole payoff of the tranche premium and default legs with one single copula (model) with parameter $\bar{\rho}_{3,6}$.

Base correlation is inconsistent at the level of single tranche: we value different parts of the same payoff with different models, i.e. part of the payoff (involving $\loss_{0,3}$) is valued with a copula in $\rho_3$, while a different part (involving $\loss_{0,6}$) of {\bf the same} payoff is valued with a copula in $\rho_6$.

We will focus on implications for base correlation later on, and deal now with compound correlation. The market data we take as inputs  are detailed in Table~\ref{tab:tableCompInvert}. In Figure~\ref{fig:comp_invert_skew} we present the compound correlation smile we imply from this set of market data. We notice that there is no bar corresponding to the 6-9\% tranche: from the market spread of the tranche we cannot imply a compound correlation. We see in the following how this problem is not atypical.

\begin{figure}[htp]
\centering
\includegraphics[width=0.45\textwidth,clip]{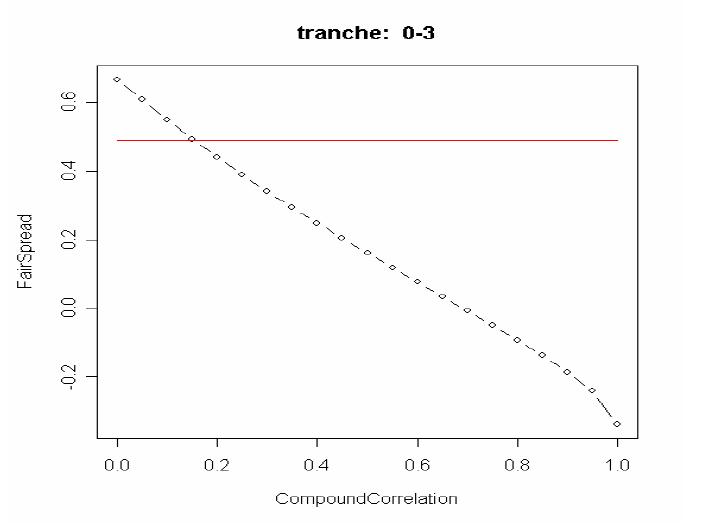}
\includegraphics[width=0.45\textwidth,clip]{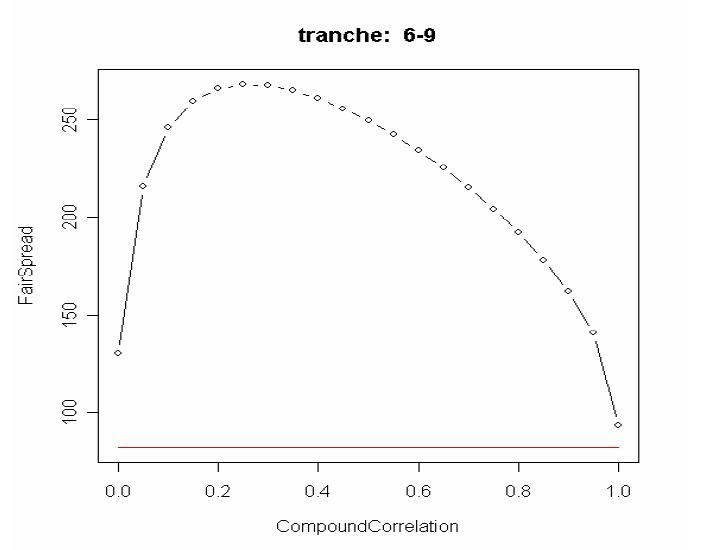}
\caption{
\label{fig:compCorrTrancheMappa}
DJi-Traxx S5 10 year Compound Correlation Invertibility. Tranche Market spread (red line) versus theorethical tranche spread obtained varying the compound correlation between 0 and 1 (dotted black line).
}
\end{figure}

We have just seen that on a particular date we cannot imply a compound correlation from the market spread of the 6-9\% tranche. To see why this is the case we investigate further this date plotting in Figure~\ref{fig:compCorrTrancheMappa} the fair market spread as a function of the compound correlation: the equity tranche is quoted upfront (0.49 means $49\%$) and all other tranches are quoted in number of running basis points (360 means $3.60\%$ per annum). The red flat line is the level of the market spread.

\begin{figure}[ht!]
\begin{center}
\includegraphics[width=0.9\textwidth]{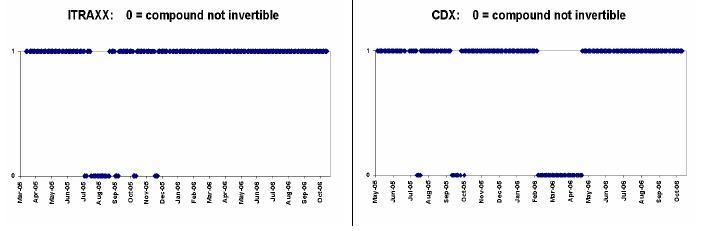}\vspace*{-0.5cm}
\end{center}
\caption{
\label{fig:invertibCompoundOld}
Compound correlation invertibility indicator (1=invertible, 0=not invertible) for the DJi-Traxx and CDX tranches. Source Torresetti et al. (2006a).
}
\end{figure}

Further, we notice that:
\begin{enumerate}
\item for certain tranches, from the unique market spread we can imply more than one compound correlation, although this does not happen in our example of Figure~\ref{fig:compCorrTrancheMappa} where the flat red line crosses the dotted black line at most in one point;
\item given a market spread we are not always guaranteed we can imply a compound correlation, as we see for example in the 6-9\% tranche of Figure~\ref{fig:compCorrTrancheMappa} (there is no intersection between the flat red line and the dotted black line).
\end{enumerate}

%
%



We have seen that on 3rd-aug-2005 we cannot imply a compound correlation for the 6-9\% tranche. In Figure~\ref{fig:invertibCompoundOld}, taken from Torresetti et al. (2006b), we see how this problem is not limited to a sporadic set of dates but is rather affecting clusters of dates.
From March 2005 to November 2006, the sample on which the analysis in Torresetti et al. (2006b) is based, the non invertibility of the compound correlation concerned for DJi-Traxx the 10 year 6-9\% tranche, and for CDX the 10 year 7-10\% tranche and, marginally, the 10 year 10-15\%.

\begin{figure}[htp]
\centering
\includegraphics[width=0.45\textwidth,clip]{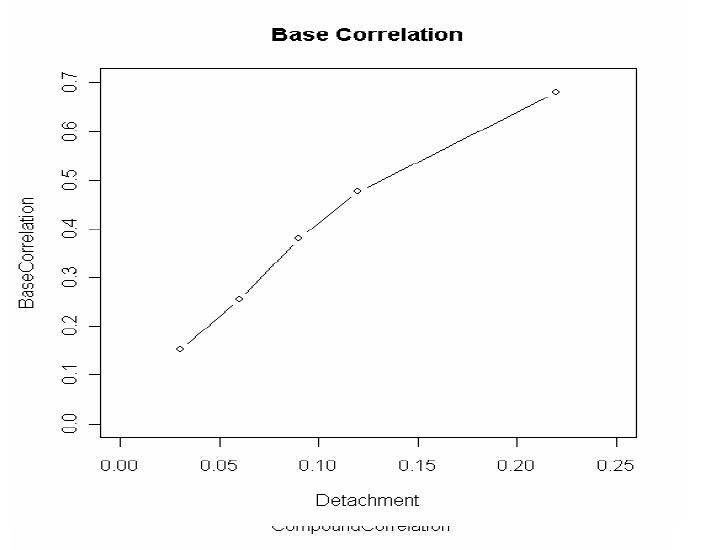}
\includegraphics[width=0.45\textwidth,clip]{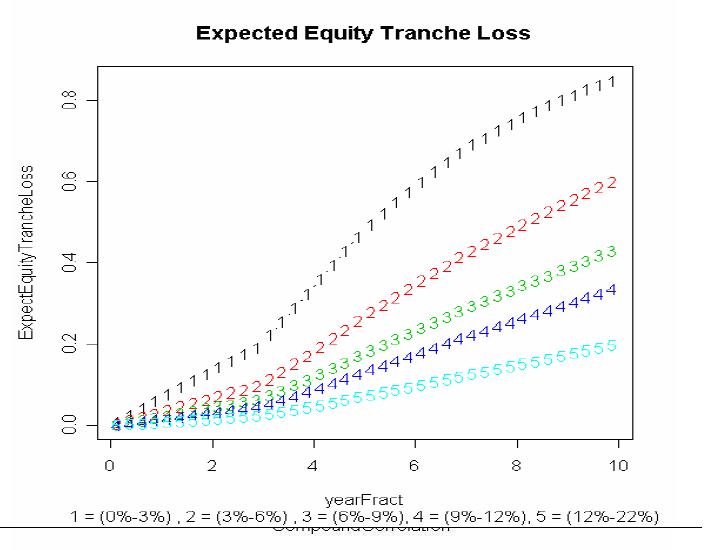}\\
\caption{
\label{fig:comp_invert_base_map}
Base Correlation calibrated to the DJi-Traxx S5 10 years tranche spreads on August, 3 2005 (left chart) and Expected Equity tranche loss (detachment set to the standardized tranches detachment) corresponding to the calibrated base correlation (right chart).
}
\end{figure}

\begin{figure}[ht!]
\centering
\includegraphics[width=0.45\textwidth,clip]{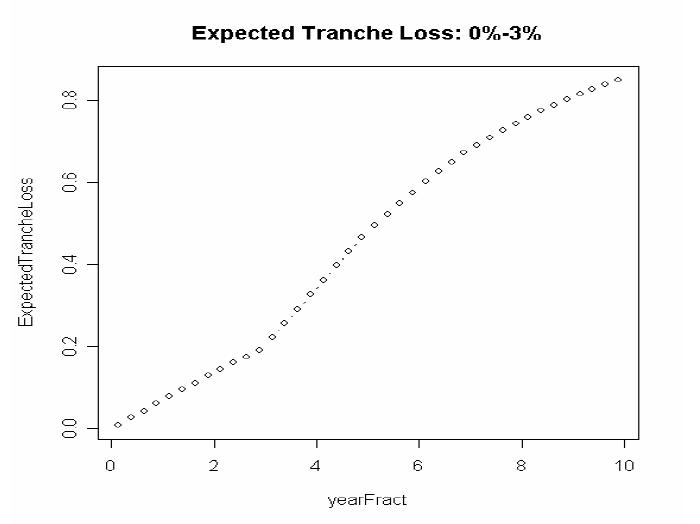}
\includegraphics[width=0.45\textwidth,clip]{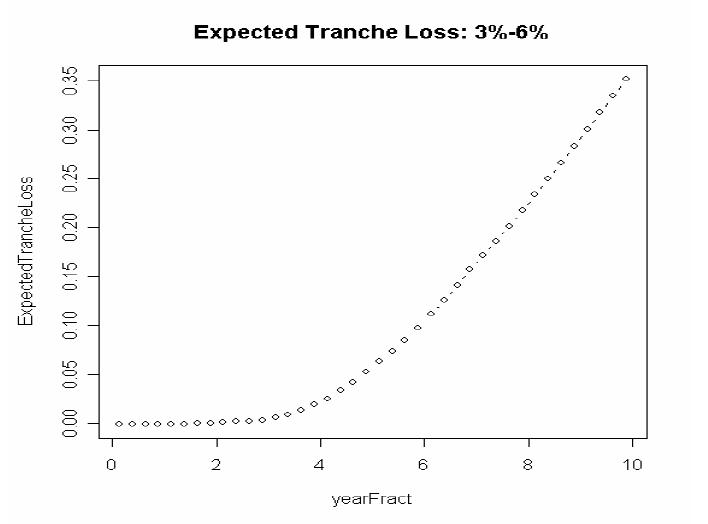}\\
\caption{
\label{fig:comp_invert_etl}
Expected tranche loss as a function of time derived from the base correlations calibrated to the DJi-Traxx S5 10 years tranche spreads on August, 3 2005.
}
\end{figure}

\subsection{Base correlation}

%
%
%
%

Here we will illustrate base correlation and we will see how compound correlation problems are overcome but at  the expense of introducing a deeper inconsistency.

%
%
%
%

In Figure~\ref{fig:comp_invert_base_map} we plot the Base Correlation calibrated to the market data in Table~\ref{tab:tableCompInvert} and the Expected Equity Tranche Loss for the various detachment points as a function of time.
\[ \Ex{0}{\lossnorm^{0,B}_t}, \ \ B=3\%,6\%, 9\%, 12\%, 22\%. \]
From these expectations, using equations (\ref{lossasdifferenceeq}), we can compute the Expected Tranche Loss $\mathbb{E}_0[\lossnorm^{A,B}_t]$, plotted in Figure~\ref{fig:comp_invert_etl} as a function of time.

From Figure~\ref{fig:comp_invert_base_map} we note that the base correlation is a much smoother function of detachments than compound correlation.   Also, to price a non-standard tranche, say a 4-15\% tranche, we can interpolate the non-standard attachment 4\% and detachment 15\% whereas with the compound correlation we do not know exactly what to interpolate (since with compound correlation there is a unique correlation associated to each tranche, i.e. correlation is associated with two points rather than a single one).

%
%
%
%


As we can see from our examples, also the base correlation approach is not immune from inconsistencies.  In fact in Figure~\ref{fig:comp_invert_etl_esempio} we note that already in 2005, taking the 6-9\% tranche as an example, the expected tranche loss becomes initially slightly negative.  This inconsistency arises from the different base correlations we use in Equation~\eqref{lossasdifferenceeq} to compute the two expected tranche loss terms in A and B.

\subsection{Summary on implied correlation}\label{sec:summary}

Is base correlation a solution to the problems of compound correlation ?

\begin{figure}[t]
\centering
\includegraphics[width=0.65\textwidth,clip]{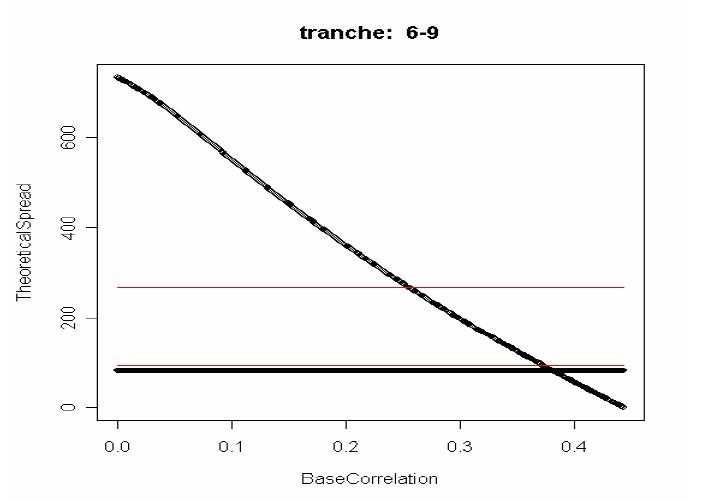}\\
\caption{
\label{fig:comp_invert_etl2}
Theoretical 10 year DJi-Traxx S9 6-9\% tranche spread as a function of the detachment base correlation keeping fixed the attachment base correlation (dotted black line), market tranche spread (flat black line) and minimum and maximum tranche spread invertible from the compound correlation (flat red lines).
}
\end{figure}

The answer is in the affirmative and this can be clearly seen for example in Figure~\ref{fig:comp_invert_etl2} where we plot the fair tranche spread as a function of the base correlation on the detachment point for each tranche, given the base correlation on the attachment point set equal to the calibrated base in the left chart of Figure~\ref{fig:comp_invert_base_map}.

This gives us an idea of the range of the tranche spread we can calibrate using base correlation. These plots of Figure~\ref{fig:comp_invert_etl2} can be compared with the plots in Figure~\ref{fig:compCorrTrancheMappa}, showing the fair tranche spread as a function of compound correlation. In Figure~\ref{fig:comp_invert_etl2} the thick black line is flat at the level of the market spread for the tranche. The two thin red lines are the minimum and maximum spread we are able to obtain by varying compound correlation.

We note that for each tranche the fair spread is a monotonic function of the base correlation on the detachment point  and also that the range of market spreads that can be attained by varying base correlation is much wider than the corresponding one for compound correlation. Consider for example the 6-9\% tranche in Figure~\ref{fig:compCorrTrancheMappa}.  The tranche spread that can be inverted in a compound correlation setting lies between 93 and 268 bps,  whereas from Figure~\ref{fig:comp_invert_etl2} the tranche spread that can be inverted in a base correlation setting lies in the wider range between 0 and 732 bps.

%

In Figure~\ref{fig:comp_invert_etl2} we did not plot the market spread for all base correlations between 0 and 1 because beyond a certain point the fair tranche spread becomes negative. Recall once again that in Equation~(\ref{lossasdifferenceeq}) we use two different correlation parameters for different parts of the same payoff: when these two correlations are very different from each other (the detachment correlation is much higher than the attachment one) the inconsistency of a negative expected tranche loss becomes more evident.


\begin{figure}[t]
\centering
\includegraphics[width=0.45\textwidth,clip]{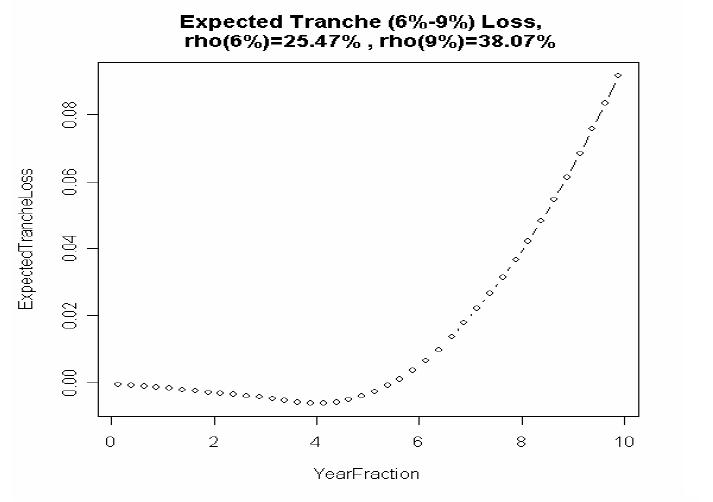}
\includegraphics[width=0.45\textwidth,clip]{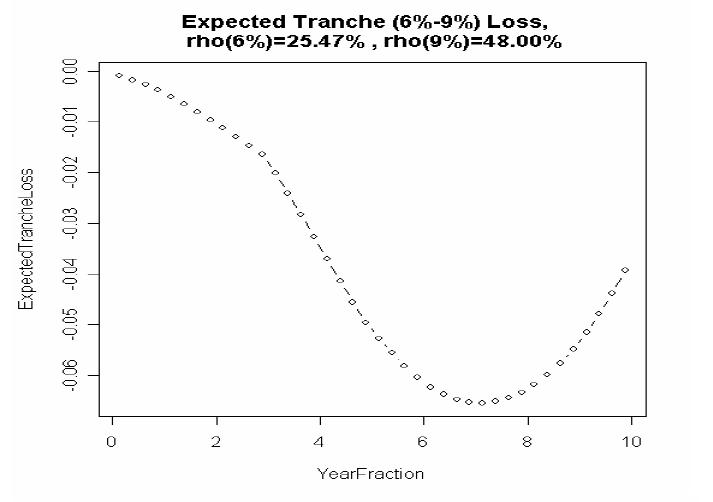}
\caption{
\label{fig:comp_invert_etl_esempio}
Expected tranche loss of the DJi-Traxx S5 10 year tranche for two different levels of the detachment base correlation: 38\% (left chart) and 48\% (right chart).
}
\end{figure}

Consider for example the 6-9\% tranche. In Figure~\ref{fig:comp_invert_etl_esempio} we plot in the abscissas the year fraction of the tranche payment dates and in the ordinates the Expected Tranche (6-9\%) Loss. For both graphs in Figure~\ref{fig:comp_invert_etl_esempio} the tranche attachment correlation is the calibrated base on the $6\%$ detachment. The tranche detachment $(9\%)$ correlation is set to the calibrated base on the left hand graph $(38.07\%)$ and to an arbitrarily high level $(48\%)$ on the right hand graph.

We can see a markedly negative profile for the expected tranched loss, which clearly violates no arbitrage constraints. Indeed, the loss must be non-negative and non-decreasing in each paths through time, and tranching does not alter that. More simply, the tranched loss at a given point in time is a non-negative random variable, and its expected value needs to be non-negative itself. This does not happen in our example, in that the basic non-negativity constraint is violated. This is a strong drawback of the base correlation paradigm at a very basic level.


We have presented notions of implied correlation centered on the Gaussian copula. Alternative copula specifications are possible. Indeed, Hull and White (2004) show that on a particular date the ``double-t copula" can consistently reproduce tranche spreads without skew in the correlation parameter.
See the full book of Brigo, Pallavicini and Torresetti (2010) for more details on the double-t copula. This is unlikely to represent a solution to the lack of consistency in calibration, given both the low number of parameters in the model and a number of numerical issues with the procedure.

There are, overall, two major problems with the use of implied correlation as a parameter in a Gaussian copula model as we described it above, even if one is willing to accept the flattening of 7750 parameters into one: first, the model parameter changes every time we change tranche, implying very different and inconsistent loss distributions on the same pool. Second, there is a total lack of dynamics in the notion of copula. We address the first problem now, showing that the financial literature has been doing that well before the current crisis started in 2007. We will address the second problem in later sections.

\section[Consistency across capital structure: implied copula]{Consistency across capital structure:\\ implied copula}\label{sec:impliedcopula}

In the implied copula approach, a factor copula structure is assumed, similarly to the One-Factor Gaussian Copula approach seen earlier. However, this time we do not model the copula explicitly, but we model default probabilities conditional on the systemic factor $S$ of the copula:  the copula will then be ``hidden" inside these conditional probabilities, that will be calibrated to the market. Hence the name ``implied" copula.
In illustrating the implied copula we will also assume a large pool homogeneous model, in that the default probabilities of single names will be taken all equal to each other and the pool of credit references is assumed to be comprised of an infinite number of credit references.

Let us consider, for simplicity, survival probabilities that are associated to a constant-in-time
hazard rate. We know that if we have a constant-in-time (possibly random) hazard rate $\lambda$ for name $i$ then the survival probability is
\[ \mathbb{Q}(\tau_i > t) = \mathbb{E}[\exp(-\lambda t)] \]

The implied copula approach postulates the following ``scenario"
distribution for the hazard rate $\lambda$ conditional on the
systemic factor $S$:
\[ \lambda|S \sim \left\{ \begin{array}{ccc} \mbox{Conditional hazard rate} & \mbox{Systemic scenario} &\ \  \mbox{Scenario probability}\\\lambda_1 & S=s_1 &
p_1\\
\lambda_2 & S=s_2 & p_2 \\ \vdots & \vdots & \vdots \\
\lambda_h & S=s_h & p_h
\end{array} \right.
\]

This way the default probability for each single name $i=1,...,M$ is, conditional on the systemic factor $S$,
\[ \mathbb{Q}(\tau_i < t|S=s_j)
= \  1-\exp(-\lambda_j t) .  \]
Compare with the Gaussian factor copula case:
\[ \mathbb{Q}(\tau_i < T |S=s_j) =
\Phi\left( \frac{\Phi^{-1}( 1 - \exp(-\Lambda(T)))-\sqrt{\rho}
s_j}{\sqrt{1-\rho}}\right)\]

Unconditionally, the implied copula yields the default probabilities

\[ \mathbb{Q}(\tau_i < t) = \sum_{j=1}^h p_j \mathbb{Q}(\tau_i < t|S=s_j)
= \sum_{j=1}^h p_j \left( 1- \exp(-\lambda_j t) \right)   \]

Conditional on $S$, all default times are independent, have the
same hazard rate and their hazard rates are given by the above
scenarios.


By resorting to the infinite pool approximation, fully illustrated in the full book by Brigo, Pallavicini and Torresetti (2010), for the Gaussian factor copula in the LHP version, we have that the conditional default indicators $1_{\{\tau_i < T | S=s_j \}}$ are i.i.d. for all $i$, so that their sample average as their number tends to infinity tends to the single common true mean:
\[
\bar{C}^M_T(s_j) := \frac{1}{M}\sum_{i=1}^M 1_{\{\tau_i < T | S=s_j \}} \rightarrow \mathbb{E} 1_{\{\tau_i < T | S=s_j \}} =
                    \mathbb{Q}\{\tau_i < T | S=s_j \} = 1 - e^{- \lambda_j T}
\]
when $M$ tends to $\infty$.

Again, this way we avoid taking expectations, except the final one with respect to $S$, since conditional on $S=s_j$ all randomness has been ruled out by the law of large numbers and both the default rate and the loss are completely determined.



We will assume, in line with the market convention, that the protection payment will be calculated on the average outstanding notional between any two protection premium payment dates.  Conditional on the systemic factor realization $S=s_j$ the premium leg tranche value will be:
\begin{align*}
\prmleg_{A,B}^j  & :=   \upfronttr_0 +  \cdstr_0   \dv_{A,B}^j  \\
\dv_{A,B}^j  & :=   \sum_{i=1}^b \, \delta_i D(0,T_i)  ( 1 - L(0.5(T_i+T_{i-1}))_{A,B}^j ) \\
L(t)_{A,B}^j & := \frac{ \min(B-A , \max( ( (1 - R_j) (1 - \exp(- \lambda_j t ) ) - A , 0 ) ) }{ B - A }
\end{align*}
where $\upfronttr_0$ and $\cdstr_0$ are the market mid upfront and running spread for the tranche A,B with maturity $T_b$.

We will discretize the loss increments, entering the calculation of the discounted default leg payoff, on the same set of dates of the discounted premium leg payoff calculation: the protection premium payment dates.  We will also assume that on average the loss increment arrives at the middle of each time interval $[T_{i-1},T_i]$.
\[
\defleg_{A,B}^j  :=   (1-R_j) \sum_{i=1}^b \, D \left( 0 , 0.5(T_i + T_{i-1}) \right) ( L(T_i)_{A,B}^j - L(T_{i-1})_{A,B}^j )
\]
where $R_j$ is a deterministic function of the probability of default conditional on the realization of the systemic factor $S$, $R_j=R ( 1- \exp ( - \lambda_j T_b ) )$ : we will expand more on this issue in the next section.

In case of the index the above definitions are still valid except for the DV01 which becomes:
\[
\dv_{0,1}^j  :=   \sum_{i=1}^b  \delta_i D(0,T_i) ( 1 - L(0.5(T_i+T_{i-1}))_{0,1}^j / (1 - R_j) )
\]
%
%
We will call $\prmleg_{A,B}$ and $\defleg_{A,B}$ the column vector that stacks respectively all discounted premium leg and default leg values  conditional on the systemic factor $s$ states:
\[
\prmleg_{A,B} = ( \prmleg_{A,B}^1,\ldots,\prmleg_{A,B}^h)^T.
\]
\[
\defleg_{A,B} = ( \defleg_{A,B}^1,\ldots,\defleg_{A,B}^h)^T.
\]
%
%
%

Then, we integrate against $S$, simply summing over all possible hazard rate scenarios multiplying by the scenario probability, to obtain the tranche unconditional price.
%
%
In matrix notation the receiver tranche value can be rewritten as
\[
\npv_{A,B}  :=  P^T ( \prmleg_{A,B}  -  \defleg_{A,B} )
\]
where $P:=(p_1,\ldots,p_h)^T$ is the column vector with the systemic factor probability distribution.

\subsection{Recovery rate}\label{sec:recoveryr}

In this respect, we mention a relationship between default
probabilities and recovery rates that may be necessary to fit the
market correlation skew in periods of turmoil (e.g July 2005).

Following results of an empirical study by Hamilton et al (2005), Hull and White suggest to change recovery in each scenario by linking it to
the conditional probability of default in that scenario:
\[  \rec_j = 0.52 - 6.9 (1-\exp(-\lambda_j 5y)) . \]

This functional form will assign a recovery 0\% to all default rate above 8\%.  Alternatively one could fit a logarithmic relationship between recovery rates and default rates:
\[  \rec_j = a  \log (1-\exp(-\lambda_j 5y)) . \]

\begin{remark}\label{rem:randrecimplcop} {\bf (Random recovery as a function of the systemic factor).} The above formula makes the recovery a function of the intensity conditional on the systemic factor. This way recovery becomes a function of the systemic factor. An approach that makes recovery a function of the systemic factor in the inconsistent base correlation framework is in Amraoui and Hitier (2008).
\end{remark}

\begin{figure}[t]
\begin{center}
\includegraphics[width=0.45\textwidth]{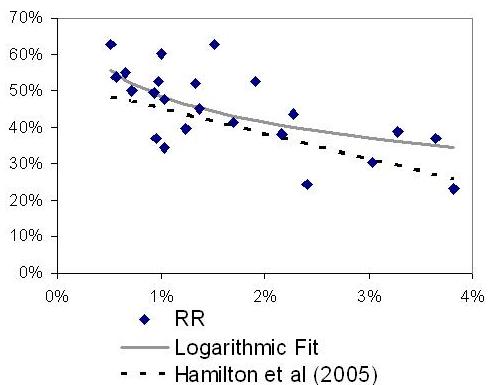}
\includegraphics[width=0.45\textwidth]{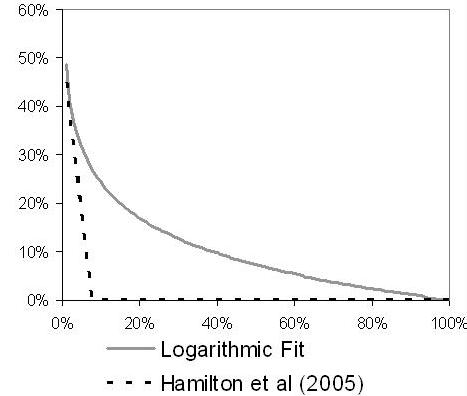}
\end{center}
\caption{
\label{fig:recovFit_inizio}
Issuer Weighted All Rated Bond Default Rate versus Issuer Weighted Recov Rate Senior Unsecured (left chart), data from 1982 to 2004 (Source: Moody's) and different functional forms between default rates and recovery rates (right chart): logarithmic (grey line) and floored linear (dotted line).
}
\end{figure}

The dots in the left panel of Figure~\ref{fig:recovFit_inizio} are the data points we have used to fit the parameter $a$: Issuer Weighted All Rated Bond Default Rate and Issuer Weighted Recovery Rate Senior Unsecured from 1982 to 2004 taken from Moody's.

From 1982 to 2004 the Issuer Weighted All Rated Bond Default Rate has been always below 3.82\%.  Therefore in the left panel of Figure~\ref{fig:recovFit_inizio} we cannot fully appreciate the different implication of the two functional forms when applied to the implied copula as we can do instead in the chart in the right panel.


%
%
%

\subsection{Calibration of implied copula}\label{sec:calibrimplcop}

When calibrating all the $5$ year DJ-iTraxx tranches we end up minimizing a constrained sum of squares. If we call $\npv$ the matrix with all tranches (columns) discounted payoff for all possible states (rows)
\[
\npv = \left[
\begin{array}{ccc}
 \npv_{0\%,3\%}^1  & \dots  &   \npv_{22\%,100\%}^1 \\
 \dots & & \\
 \npv_{0\%,3\%}^h  & \dots  &   \npv_{22\%,100\%}^h
\end{array}
\right]
\]
then, calibration becomes simply
\begin{equation}
\label{eq:implCopHWnoC}
 \underset{P = (p_1,\dots,p_h)^T}{\operatorname{argmin}} P^T \ \npv \ \npv ^T \ P \\
\end{equation}
subject to:
\[
 \sum_{i=1}^h p_i = 1 \;,\quad p_i \geq 0  \ \ \  i=1,\dots,h \\
\]

Note that for the base correlation calibration the market tranches had to have consecutive adjacent attachment and detachment points spanning the whole capital structure; if that was not the case some base correlation interpolation assumption had to be introduced. With the Implied Copula this is not necessary in that the tranches to be calibrated need not span the whole capital structure or be adjacent.


In the implied copula framework, Hull and White (2006) calibrate the scenario probabilities $p$ while pre-assigning the hazard rate scenarios $\lambda_j$ exogenously.   The number of hazard rate scenarios $\lambda_j$ can be seen empirically to be quite large, up to $h=30$, in order to be able to fit market data with a good precision.  In this case the above optimization has too many degrees of freedom which might result in a very good fit but quite an irregular scenario probability distribution.
In order to cope with this problem Hull and White (2006) propose to add to the target function a quantity
that penalizes changes in convexity in the patterns of the scenario probabilities plotted against the default probabilities
associated to each scenario.

Torresetti et al. (2006c) propose a different approach with respect to Hull and White (2006) where there is no need to select a regularization coefficient. The main differences between the two approaches are summarized as follows.
\begin{itemize}
\item  Assign 125 possible states to the system factor $S=s_j$.  The hazard rates associated to each state are this time such that the pool default rate at maturity is equal to $(j-1)/125$, where $j=1,\ldots,125$.
\item  In our version perform a two stage optimization that will assure that all tranches are priced within the bid-ask spread without the need to choose a regularization coefficient.
\end{itemize}


The numerical solution 
of the fist stage optimization, i.e. the solution to the problem set out in \eqref{eq:implCopHWnoC}, can be expressed as a deviation of each tranche theoretical spread from the market mid quote.
%
If the standardized mispricing is between $-1$ and $+1$ then the vector of systemic states probabilities $P$ results in a calibration of the theoretical tranche spread $\npv_{A,B} / \dv_{A,B}$ that lies within the bid ask.
%
%
The solution to the first stage optimization gives a numerical starting point to the second stage optimization problem, where the second differential of the state probabilities is minimized while the standardized mispricing is kept within the $[-1,1]$ interval.





\begin{figure}[t]
\begin{center}
\includegraphics[width=0.5\textwidth]{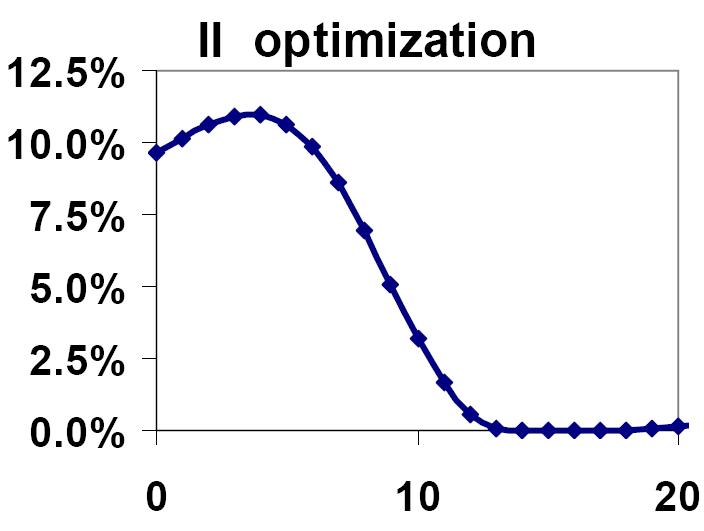}\hspace*{0.5cm}
\includegraphics[width=0.45\textwidth]{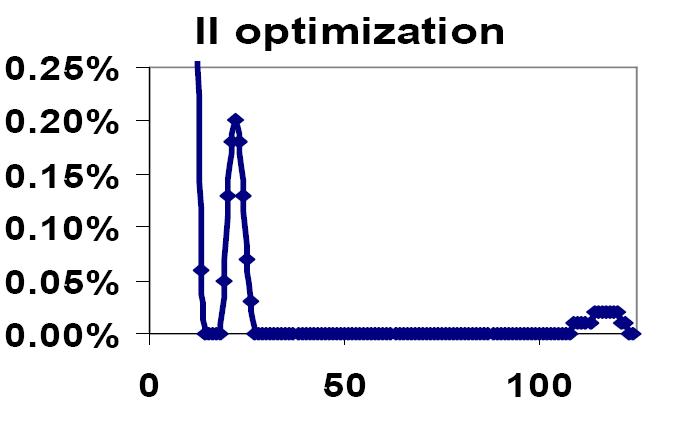}\vspace*{-1cm}
\end{center}
\caption{
\label{fig:ImplCopu_2}
Optimal pool default counting $\defrate_{5y}$ distribution resulting from the two stage optimization: distribution tail magnified in the chart on the right panel.}
\end{figure}

We consider the market quotes in Table~\ref{tab:implCopQuote} for the CDX.NA.IG 5 year tranches on June, 6 2006, while in Figure~\ref{fig:ImplCopu_2} we plot the optimal distribution resulting from the two stage optimization as outlined in Torresetti et al. (2006c) .

\begin{table}[t]
\centering
\begin{tabular}{cccc}
\hline \hline
        &  Mid        &   Bid       &  Ask    \\
\hline
0-3\%   &  32.40\%    &  31.50\%    &   33.60\%    \\
3-7\%   &  106.5 bps  &  100 bps    &   113 bps    \\
7-10\%  &  23.5 bps   &  22 bps     &   25 bps    \\
10-15\% &  10 bps     &  9.3 bps    &   10.7 bps    \\
15-30\% &  5.5 bps    &  5.1 bps    &   5.9 bps    \\
\hline
\end{tabular}
\caption{\label{tab:implCopQuote}CDX.NA.IG 5 year tranches quotes on the June, 6 2006}
\end{table}

%

\subsection{Summary of considerations around implied copula}

The implied copula as we presented it above can solve the problem of consistency with tranche quotes for a single maturity: there is a unique implied loss distribution that is consistent with all the available tranche quotes for a single maturity.
Suppose one is momentarily willing to leave aside the problem of single names modeling and hedges, which is neglected by the homogeneous pool assumptions. When CDO tranche quotes for more than one maturity are available, we cannot account consistently for them with an implied copula model. We need a arbitrage-free dynamic loss model, implying future loss distributions for the loss at different times that are consistent with the tranche quotes at each maturity.

An explicit model implying a dynamics for dependence across defaults, absent in the copula case, is the above mentioned GPL model given in Brigo Pallavicini and Torresetti (2006a,b), and to that we turn in Section~\ref{sec:gplmodel} below, leaving aside the single-name consistent but combinatorially complex GPCL model in Brigo et al (2007). Finally, we note that the loss distribution of the pool under the risk neutral measure, which is being discussed in this paper, is different from the actual loss distribution in the objective measure, as pointed out for example in Torresetti, Brigo and Pallavicini (2006c), who find the well-known presence of a risk premium. See also the discussion in Torresetti and Pallavicini (2007) concerning CPDOs, where the difference in the modes of the tail loss distribution is also visible.

Before turning to the fully dynamic loss model in Section~\ref{sec:gplmodel}, we explore en passant a model-free extraction of expected loss information from CDO quotes that can be occasionally helpful in interpolating or checking arbitrage constraints.

\section{Consistency across capital structure and maturities: the Expected Tranche Loss surface}\label{sec:eetl}

In this section we approach the issue of consistently representing the information embedded in tranches across attachments and maturities in a model-free way. Let us consider again Equation~\eqref{eq:tranche}.
%
%
\[
\cdstr_0 = \frac{ \int_0^T D(0,t)\ d\ \Ex{0}{\losstr_t} - \upfronttr_0}
                { \sum_{i=1}^b \, \delta_i D(0,T_i) (1- \Ex{0}{\losstr_ {T_i}} ) }
\]

The numerator and denominator of the above formula depend linearly on the expected tranche loss ${\mathbb E}_0[\losstr_{T_i}]$ (ETL)\footnote{Let us stress again that the actual outstanding notional in each period would be a daily average over $[T_{i-1}, T_i]$, but for simplicity we replace it with $1- \losstr_ {T_i}$, the value of the tranche outstanding notional at $T_i$.} at different times. If a term structure of tranche upfronts and spreads $\cdstr_{0,T}$ for different maturities $T$ is given, then it is possible to strip back the expectations ${\mathbb E}_0[\losstr_{T_i}]$ in a model independent way. We will address this issue in this section.

We first notice the nested structure of the expected tranche losses in time. Indeed, the NPV calculations of two tranches with the same attachment and detachment points ($A,B$) but different maturities $T_1$ and $T_2$, depend both on the expected tranche loss ${\mathbb E}_0[\losstr_{T_i}]$ for each $T_i \leq \min\{T_1,T_2\}$. For example in the NPV calculations of a 5 year maturity 0\%-3\% tranche and a 10 year maturity 0\%-3\% tranche the quantity ${\mathbb E}_0[\losstr_{T_i}]$ should be the same for the calculation of the 5 and 10 year tranches for all $T_i<5$y.

\begin{figure}[t]
\begin{center}
\includegraphics[width=0.65\textwidth]{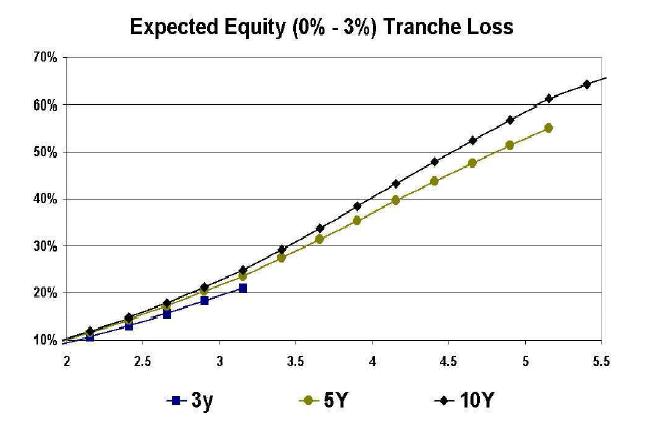}\vspace*{-1cm}
\end{center}
\caption{
\label{fig:ETL_walker}
Expected 0\%-3\% tranche loss resulting from the base correlation, One-Factor Homogeneous Finite Pool Gaussian Copula, calibration to the 3, 5 and 10 year 0\%-3\% CDX tranches on April, 26th 2006.}
\end{figure}

In the copula framework unfortunately we are not guaranteed this to be the case, regardless of the particular copula model considered, Gaussian versus Double-t, or Homogeneous versus Heterogeneous. This has to do with the copula being a static object. As an illustration, in Figure~\ref{fig:ETL_walker} we plot the expected 0\%-3\% tranche loss resulting from the base correlation, One-Factor Homogeneous Finite Pool Gaussian Copula, calibrated to the 3, 5 and 10 year 0\%-3\% tranches on January, 30th 2009.

We note that the expected 0\%-3\% tranche loss resulting from the calibration to the 3, 5 and 10 year 0\%-3\% tranches do not overlap. Thus when valuing the same expected tranche loss ${\mathbb E}_0[\lossnorm_ {T_i}^{0\%,3\%}]$ for $T_i<3$y we are using three different numbers depending on the tranche maturity even though the pool of underlying credit references is the same.

In the following we will see the details of a non-parametric model-free estimation of the basic quantities entering linearly the tranche NPV calculation, namely the expected tranche losses ${\mathbb E}_0[\losstr_{T_i}]$, which is consistent through maturities and attachments.

\subsection{Index and tranche NPV as a function of ETL}

Let us assume we are given the market spreads of a set of tranches spanning the entire capital structure. More in particular, let us assume that we have the market spreads of a set of $k$ tranches with attachments $A_j$ and detachments $B_j$ with $j=1,\ldots,k$ where $A_1=0$, $B_k=1$ and $A_{i+1}=B_{i}$ for $i=1,\ldots,k-1$.

Let us consider the standardized DJ-iTraxx tranches with detachments (3\%, 6\%, 9\%, 12\%, 22\%): all except the super-senior 22-100\%\footnote{At the time when Walker (2006) and Torresetti et al. (2006a) ran their analysis the super-senior tranche\index{Super-Senior Tranche} was just beginning to have some liquidity so that it was not considered in their historical calibration analysis of the expected equity tranche loss surface.  In fact before the end of 2005 the liquidity of the super-senior tranche was quite limited.}.
%
%
Our goal is to reprice the 5y tranches and the index: a total of 6 market quotes. To achieve this goal, given a deterministic recovery, we will be looking for the 6 unknown 5 year ETLs
that set the NPV of the instruments as close as possible to zero when the quoted spreads and upfront payments are put into the premium leg.

To compute the NPV of the tranches and index we also need the ETL on all payment dates with maturity $T_i$ shorter than 5 year: these will be obtained
interpolating for each tranche between the initial time 0, where by definition the ETL will be zero\footnote{In case the pool of underlying credit references already suffered losses, then a new version of the index is rolled out.  The existing positions in both the index and the tranches are usually promptly rolled in the new version.  Tranches of the newly rolled version see the standardized attachment and detachment adjusted to reflect the loss in subordination and the reduction in the pool outstanding balance.}, and the six 5 year unknown ETL that are being sought: 3\%, 6\%, 9\%, 12\%, 22\% and 100\%. Therefore,  once we set a possible value for the unknown 5 year ETLs in the iteration implementing the optimization, through interpolation we also set a value for all the ETL path up to 5 years that is consistent with the selected 5 years value being iterated.

Once the 5 year nodal ETLs matching the 5 years data are found, to price the 10 years tranches and index we will need also the ETL between 5 and 10 years. As before, for each tranche this will be obtained by interpolation between the expected tranche loss at 5 years and the six unknown ETLs at 10 year to be found. 

We call $f(t,h,k)$ the ETL at time $t$ of the tranche with attachment $h$ and detachment $k$. To simplify the notation we will often identify the seniority of the tranche in the capital structure of the CDO only through the detachment point $k$, writing $f(t,h,k)=f(t,k)$ when $h$ is clear from the context given the adjacent attachment and detachment points.
\[
f(t,A,B) := \Ex{0}{\lossnorm_{t}^{A,B}} .
\]

By assuming interest rates and default times to be independent and discretizing the integrals entering the NPV of the premium and default leg in Equation~\eqref{eq:tranche} on the tranche payment dates we obtain
\begin{equation}
\label{eq:SpreadTrancheF}
S_{T_b}^{A,B} = \frac{ \sum_{i=1}^b  D(0,T_i) (f(T_i,A,B) - f(T_{i-1},A,B)) }
                { \sum_{i=1}^b \, \delta_i D(0,T_i) (1 - f(T_i,A,B) ) }
\end{equation}
Then, since the tranches are adjacent, the expected portfolio loss is the summation of the ETL multiplied by the tranche depth (detachment minus attachment), namely:
\begin{eqnarray*}\label{eq:expLoss}
    f(t,0,100\%) = \sum_{i=1}^k \mathbb{E}[ \lossnorm_t^{A_i, B_i}] \left(  B_i - A_i  \right) =\mathbb{E}[ \lossnorm_t]
\end{eqnarray*}

Once given the expected portfolio loss $\mathbb{E}[\lossnorm_t]$ and the recovery rate $R$, we can compute the expected portfolio default rate as
$\mathbb{E}[\lossnorm_t]/(1-R) = f(t,0,100\%)/(1-R)$.
Hence, the index spread becomes:
\begin{equation}
\label{eq:SpreadIndexF}
S_{T_b} = \frac{ \sum_{i=1}^b  D(0,T_{i}) (f(T_i,0,100\%) - f(T_{i-1},0,100\%)) }
                { \sum_{i=1}^b \, \delta_i D(0,T_i) (1 - f(T_i,0,100\%) / (1-R) ) }
\end{equation}

We will seek to calibrate tranches (the entire capital structure except the super-senior tranche) and indices for the 3, 5, 7 and 10 year maturities in terms of all the unknown $f(t,A,B)$ terms.
%
%
The set of $4 \cdot 10 \cdot 6 = 240$  $f$s (one for each maturity, quarterly payment date and for each tranche) created by interpolating the $4 \cdot 6 =24$ basic nodal $f(t,k)$s (one for each maturity and for each detachment) will be used to set the NPV of the $24 = 4 \cdot (5 + 1)$ instruments (5 tranches and 1 index for each maturity) as close to zero as possible whilst
maintaining the following constraints




\begin{eqnarray}
	\label{eq:constrF}
 	\underset{ \{ f(3y,3\%), ... ,f(10y,100\%) \} }{\operatorname{argmin}}
  	 \sum_{T}\sum_{(A,B)}{ \mbox{M{\tiny ispr}S{\tiny tdz}}_{T,A,B}^2 }
\end{eqnarray}
subject to:
\begin{equation}\label{eq:constrF2nd}
    \left\{
        \begin{array}{l}
            0 \leq f(t,k) \leq 1 \\
            f(t_i,k) \ge f(t_{i-1},k) \\
            f(t,k_{j-1}) \ge f(t,k_{j}) \\
        \end{array}
    \right.\
\end{equation}
where
\begin{equation}\label{misprBA}
	\mbox{M{\tiny ispr}S{\tiny tdz}}_{T,A,B} = \frac{ S_{T_b}^{A,B} - S_{T_b}^{A,B,\mbox{mid}}  }{ ( S_{T_b}^{A,B,\mbox{ask}} - S_{T_b}^{A,B,\mbox{bid}} ) / 2 }
\end{equation}
and the double summation in the objective function
is taken with respect to all four maturities (3, 5, 7 and 10 year) and all available instruments, tranches plus the index.
%
Note that the ETLs $f(t,k)$ are entering the equation of the standardized mispricing~(\ref{misprBA}) of the tranches and indices 
via equations~\ref{eq:SpreadTrancheF} and~\ref{eq:SpreadIndexF} respectively.

\begin{figure}[htp]
\centering
\includegraphics[totalheight=0.3\textheight,width=0.45\textwidth,clip]{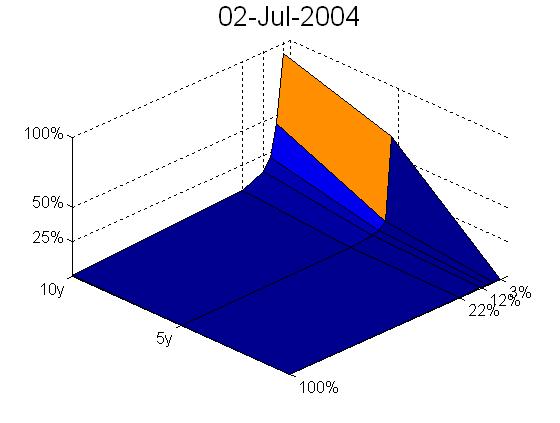}
\includegraphics[totalheight=0.3\textheight,width=0.45\textwidth,clip]{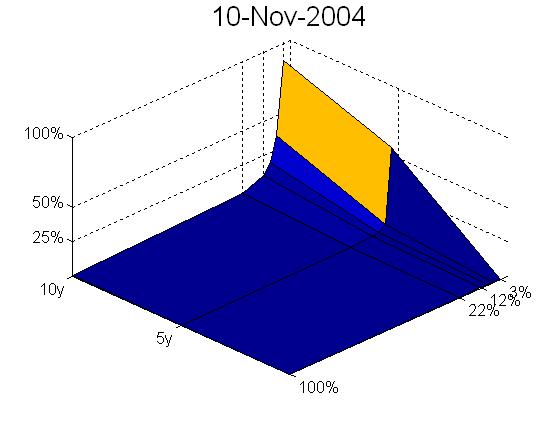} \\
\includegraphics[totalheight=0.3\textheight,width=0.45\textwidth,clip]{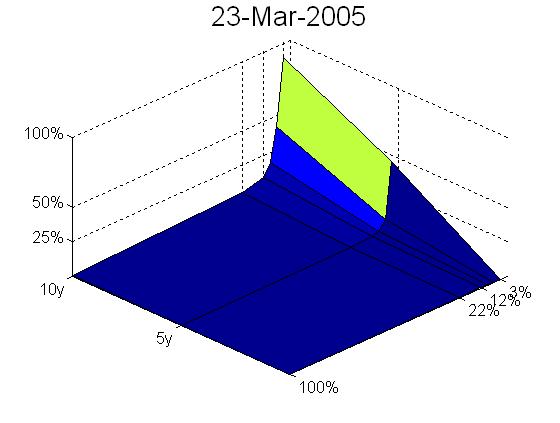}
\includegraphics[totalheight=0.3\textheight,width=0.45\textwidth,clip]{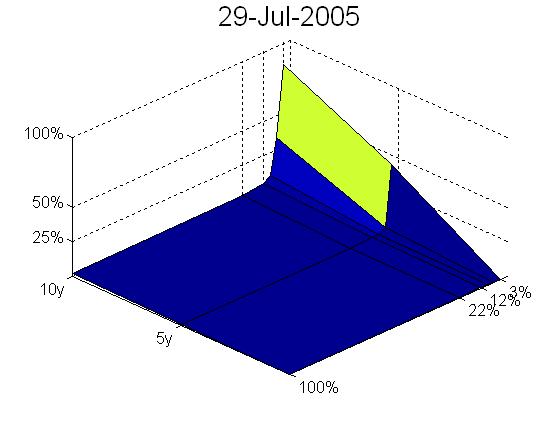} \\
\includegraphics[totalheight=0.3\textheight,width=0.45\textwidth,clip]{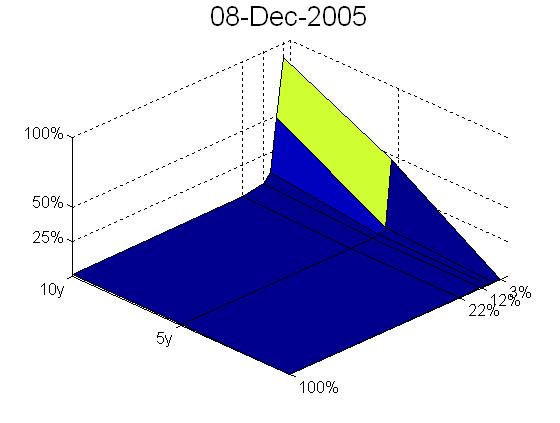}
\includegraphics[totalheight=0.3\textheight,width=0.45\textwidth,clip]{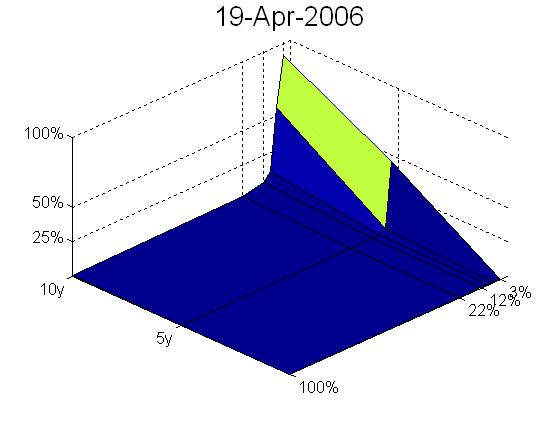} \\
\caption{
\label{fig:etlHist}
Expected Tranche Loss calibrated to the Dj-iTraxx 5 year and 10 year tranches plus index.  }
\end{figure}

In Figure~\ref{fig:etlHist} we plot the ETL $f(t,k)$ for a series of dates.  We note that the surface is not particularly smooth. To price a tranche on the same pool of credit references with non standard attachment and detachment $A$ and $B$ we would need to obtain the expected equity tranche losses (EETL) $g(T,B) = B f(T,0,B)$ and $g(T,A)= A f(T,0,A)$. EETL give a condition of no-arbitrage corresponding to requiring that the second derivative of $g(T,A)$ wrt A be non-positive, since this would be the opposite of the loss density computed in $A$, as once can see by using a Breeden Litzenberger (1978) approach.

\subsection{Numerical Results}

Our sample pre-crisis goes from November 13, 2003 to June, 14 2006 for the CDX and from June 21, 2004 to May, 23 2006 for DJ-iTraxx. From Table~\ref{tab:percMispr} we note that, except for the Dj-iTraxx pool with a linear interpolation, in all other cases we find a solution where the
theoretical spread exceeds the bid ask spread by less than one fifth ($0.4/2$) the bid ask range. In the case of the Dj-iTraxx pool with a linear interpolation we find only one date where all instruments cannot be priced within the bid ask range: in this case the theoretical spread is outside the bid ask spread by {\bf less} than one third ($0.6/2$) the bid ask range.

In all cases we find only few dates where we could not calibrate an EETL surface within the bid-ask spread of each tranche plus the index. The few dates where market instruments could not be priced exactly show a standardized mispricing that is extremely small: the highest mispricing was obtained for the DJ-iTraxx using a linear interpolation: the mispricing being only 0.3 times the bid-ask spread: $0.3 = (1.6 - 1) / 2$.


\begin{table}
\begin{center}
\begin{tabular}{ccccc}\hline\hline
                 & \multicolumn{ 2}{c}{{\bf CDX}} & \multicolumn{ 2}{c}{{\bf DJ-iTraxx}} \\\hline
    {\bf Interpolation}   & {\bf linear} & {\bf spline} & {\bf linear} & {\bf spline} \\\hline
    {\bf Number of dates} &          616 &        616 &        473 &        473 \\\hline
    {\bf \% $ \mbox{Mispr}_{\mbox{\tiny BidAsk}} > 1 $ }   &      1.0\% &      2.6\% &      0.2\% &      1.3\% \\
    {\bf \% $ \mbox{Mispr}_{\mbox{\tiny BidAsk}} > 1.2 $ } &      0.8\% &      0.2\% &      0.2\% &      0.2\% \\
    {\bf \% $ \mbox{Mispr}_{\mbox{\tiny BidAsk}} > 1.4 $ } &      0.0\% &      0.0\% &      0.2\% &      0.0\% \\
    {\bf \% $ \mbox{Mispr}_{\mbox{\tiny BidAsk}} > 1.6 $ } &      0.0\% &      0.0\% &      0.0\% &      0.0\% \\\hline
\end{tabular}
\end{center}
\caption{Percentage of sample repriced outside the bid-ask range. Sample data for CDX range from November, 13 2003 to June, 14 2006, while for DJ-iTraxx from June 21, 2004 to May, 23 2006.
\label{tab:percMispr}}
\end{table}

%

\subsection{Summary on Expected (Equity) Tranche Loss}

The notion of ETL surface across maturity and tranche attachments is as close as a model independent notion of implied dependence as possible, since it focuses on one of the most direct market objects embedded in market quotes.

Rather than going through implied correlation, based on the arbitrary assumption of a Gaussian Copula connecting defaults across names and leading to inconsistencies in the temporal axis, one considers directly quantities entering the valuation formula and implies them from market quotes given minimal interpolation assumptions. To make sure that interpolation does not interfere excessively we carry out the calibration through two different
interpolation techniques (linear and splines).

While in our framework the bid/asks of the instruments enter the target function we aim at minimizing in order to imply the surface, in Walker's (2006) framework the instruments NPVs must be exactly zero. By including bid/asks as we do the no-arbitrage constraints are satisfied across the vast majority of dates, in particular we find less violations of the no-arbitrage condition than in Walker's (2006).\index{ETL!consistency}

The method appears to be helpful as a first model-independent procedure to deduce implied expected loss surfaces from market data, allowing one to check basic no-arbitrage constraints in the market quotes. It is also of immediate use to value tranches with nonstandard attachments and maturities and forward starting tranches, although excessive {\it extrapolation} is to be avoided.

\section[A fully consistent dynamical model: the Generalized-Poisson Loss model]{A fully consistent dynamical model:\\ the Generalized-Poisson Loss model}\label{sec:gplmodel}

We finally turn to a full fledged arbitrage free and consistent dynamic loss model, improving both on the implied copula through time consistency and also on ETL through real dynamics.
The Generalized Poisson Loss (GPL) model can be formulated as follows. Consider a probability space supporting a number $n$ of independent Poisson processes $N_1, \ldots, N_n$ with time-varying, and possibly stochastic, intensities $\lambda_1, \ldots, \lambda_n$ under the risk neutral measure $\mathbb{Q}$. These intensities $\lambda_j$ are intensities of defaults of sectors generating losses of size $\alpha_j$, and should not be confused with the single name default intensities introduced earlier.

The risk neutral expectation conditional on the market information up to time $t$, including the pool loss evolution  up to $t$, is denoted by $\mathbb{E}_t$. Intensities, if stochastic, are assumed to be adapted to such information. Define the stochastic process
\begin{equation}
\label{eq:gpldriver}
\gpl_t := \sum_{j=1}^n \alpha_j N_j(t),
\end{equation}
for positive integers $\alpha_1,\ldots,\alpha_n$. In the following we refer to the $\gpl_t$ process simply as the GPL process. This process may be considered as the driving process either for the cumulated portfolio loss $\lossnorm_t$ or the default counting process $\defratenorm_t$.

\subsection{Loss dynamics}

The underlying GPL process $\gpl_t$ is non-decreasing and takes arbitrarily large values, given large enough times. The actual portfolio cumulated loss and the re-scaled number of defaults processes are non-decreasing, but limited to the interval $[0,1]$. Thus, by following Brigo, Pallavicini and Torresetti (2006a,b), we consider the deterministic non-decreasing function $\Psi:\mathbb{N}\cup\{0\}\rightarrow[0,1]$ and we define the cumulated portfolio loss process $\lossnorm_t$ as
\begin{equation}
\label{eq:losseq}
\loss_t := \Psi_\loss(\gpl_t) := \min(\gpl_t,M')
,\quad {\rm and} \quad
\lossnorm_t := \frac{\loss_t}{M'}
\end{equation}
where $1/M'$, with $M'\ge M>0$, is the minimum jump for the loss process.  For example we could choose $M'$ such that $1/M'=(1-0.4)/125$.  In case we want to allow for a finer resolution of the loss we could choose instead $M'$ such that $1/M'=(1-0.4)/(2 \cdots 125)$.

The related density (defined on integer values since the law is discrete) is
\[
p_{\loss_t}(x) = p_{\gpl_t}(x) \ind{x<M'} + \mathbb{Q}\{Z_t \ge M'\} \ind{x = M'}
\]
where the marginal distribution $p_{\gpl_t}$ of the process can be directly computed by means of an inverse Fourier transform starting from the characteristic function $\varphi_{\gpl_t}$, which can be explicitly calculated for some relevant choices of Poisson cumulated intensities distributions, and it is given by
\begin{equation*}
\varphi_{\gpl_t}(u) = \Ex{0}{\exp(i u \gpl_t)} = \Ex{0}{ \,\exp\left(\sum_{j=1}^n \Lambda_j(t)(e^{iu\alpha_j}-1)\right) }
\end{equation*}

Also the intensity of $\loss_t$, i.e. the density of the absolutely continuous compensator of $\loss_t$ (see for example Giesecke and Goldberg (2005)), can be computed directly and is given by
\begin{equation*}
h_\loss(t) = \sum_{j=1}^n \min(\alpha_j,(M'-\gpl_{t^-})^+)\lambda_j(t),
\end{equation*}
The intensity $h$ goes to zero when $Z$ exceeds $M'$, which corresponds to total loss, as expected. Further, if all the possible integer jump sizes between 1 and $M'$  are allowed, i.e. if $\alpha_j=j$ and $n=M'$, the intensity $h_\loss$ jumps whenever the cumulated portfolio loss process $\loss$ jumps.

\begin{remark}
The intensity jumps downwards, and this would seem to go in the opposite direction with respect to self-excitedness, which is considered a desirable feature of loss models in general. However, self-exciting features are embedded in our model, and they are embedded in the possibility to have several defaults in small intervals, contrary to most approaches to loss modeling.
\end{remark}

\begin{remark}
Since in the present paper we focus only on the calibration of CDO tranches, which depend only on the loss marginal distribution, we address the readers to Brigo, Pallavicini and Torresetti (2006a,b) for an extensive analysis of candidate  spread and recovery dynamics, and to Brigo, Pallavicini and Torresetti (2007) for further discussion, including consistency with single names data, see also the full book by Brigo, Pallavicini and Torresetti (2010). We point out that significant progress and testing in loss modeling will be possible only when more liquid market quotes for tranche options and forward start tranches will be available.
\end{remark}

\subsection{Model calibration}

Since credit indices and CDO tranches depend only on the expectation of the portfolio cumulated tranched loss ($\lossnorm_t$) and of the re-scaled number of defaults ($\defratenorm_t$), we avoid to directly introduce stochasticity on the process intensities.
We work with the basic GPL model specification given by the driving GPL process $Z$ in \eqref{eq:gpldriver}, which we use in modeling the pool loss through \eqref{eq:losseq}. In this basic formulation each Poisson mode $N_j$ has a
deterministic piecewise-constant intensity $\lambda_j(t)$.

Given that we have modeled the pool loss $\lossnorm_t$ directly, we do not characterize completely the re-scaled default counting process $\defratenorm_t$, but we give only its expectations. Indeed, such expectations are the only information on default counting that are implicit in Index market quotes~(\ref{eq:index}), whereas tranche quotes (\ref{eq:tranche}) depend only on the loss and not on default counting explicitly. We thus assume:
\begin{equation*}
\Ex{0}{\defratenorm_t} := \frac{1}{1-{\cal R}} \,\Ex{0}{\lossnorm_t}
\quad {\rm with} \quad
0 \le {\cal R} < 1- \Ex{0}{\lossnorm_{T_b}}
\end{equation*}
where the range of definition of the constant ${\cal R}$ is taken in order to ensure that at each time $t$ the expected value of the re-scaled number of defaults is greater, or equal to, the cumulated portfolio loss, and that both be smaller or equal to one. The parameters calibrated in this version of the model are the $\alpha$s and the $\lambda$s. A detailed model calibration can be found in Brigo, Pallavicini and Torresetti (2006a).

\subsection{Calibration results}

The GPL model is calibrated to the market quotes observed weekly from May 6, 2005 to October 18, 2005. We take ${\cal R}=30\%$, following Albanese et al (2006), as reference value for the recovery rate in the DJi-TRAXX Europe market for spot and forward contracts. The quality of our calibration below is not altered if we select a value ${\cal R}=40\%$ resembling the recovery typically used in simplified quoting mechanisms in the market, see Brigo, Pallavicini and Torresetti (2006a,b) and (2007) for some examples. 

\begin{table}[t]
\begin{center}\small
\begin{tabular}{|c|c|*4{>{\centering\arraybackslash}p{.62cm}>{\centering\arraybackslash}p{0.42cm}>{\centering\arraybackslash}p{0.52cm}|}}\hline
 & \multicolumn{1}{c}{{\bf Maturities}} & \multicolumn{3}{c}{3y} & \multicolumn{3}{c}{5y} & \multicolumn{3}{c}{7y} & \multicolumn{3}{c|}{10y} \\\hline
 & {\bf Att-Det} & $\Pi$ & $\Delta$ & $E$ & $\Pi$ & $\Delta$ & $E$ & $\Pi$ & $\Delta$ & $E$ & $\Pi$ & $\Delta$ & $E$ \\\hline
{\bf Index}   &            &   38 &   4 & 0 &   54 &   1 &  0 &   65 &   3 &  1 &   77 &   2 &   0 \\\hline
{\bf Tranche} &        0-3 & 2060 & 100 & 1 & 4262 & 118 &  8 & 5421 & 384 & 73 & 6489 & 124 & -21 \\
              &        3-6 &   72 &  10 & 0 &  173 &  68 &  0 &  398 &  40 & -8 &  590 &  20 &   1 \\
              &        6-9 &   28 &   6 & 0 &   57 &   6 &  0 &  141 &  17 & -5 &  188 &  15 &   2 \\
              &       9-12 &   13 &   2 & 0 &   31 &   5 &  1 &   72 &  20 & -3 &   87 &  15 &   6 \\
              &      12-22 &    3 &   1 & 0 &   21 &   3 &  0 &   42 &  13 & -3 &   60 &  10 &  -3 \\\hline
\end{tabular}
\end{center}
\begin{center}\small
\begin{tabular}{|c|c|*4{>{\centering\arraybackslash}p{.62cm}>{\centering\arraybackslash}p{0.42cm}>{\centering\arraybackslash}p{0.52cm}|}}\hline
 & \multicolumn{1}{c}{{\bf Maturities}} & \multicolumn{3}{c}{3y} & \multicolumn{3}{c}{5y} & \multicolumn{3}{c}{7y} & \multicolumn{3}{c|}{10y} \\\hline
 & {\bf Att-Det} & $\Pi$ & $\Delta$ & $E$ & $\Pi$ & $\Delta$ & $E$ & $\Pi$ & $\Delta$ & $E$ & $\Pi$ & $\Delta$ & $E$ \\\hline
{\bf Index}   &            &  23 &  2 &  0 &   38 &  1 & 0 &   47 &  1 &   0 &   58 &  1 &   0 \\\hline
{\bf Tranche} &        0-3 & 762 & 26 & -2 & 3137 & 26 & 2 & 4862 & 76 & \underline{-89} & 5862 & 74 & \underline{157} \\
              &        3-6 &  20 & 10 &  1 &   95 &  1 & 0 &  200 &  3 &   1 &  515 & 10 & \underline{-10} \\
              &        6-9 &   7 &  6 &  0 &   28 &  1 & 0 &   43 &  2 &   1 &  100 &  4 &   3 \\
            & 9-12 & \multicolumn{3}{c|}{} &   12 &  2 & 1 &   27 &  4 &  -3 &   54 &  5 &  -4 \\
           & 12-22 & \multicolumn{3}{c|}{} &    7 &  1 & 0 &   13 &  2 &   0 &   23 &  3 &   0 \\\hline
\end{tabular}
\end{center}
\caption{\label{gpl:precrisis}
DJi-TRAXX index and tranche quotes ($\Pi$) in basis points on May 13, 2005 (upper panel) and on October 11, 2005 (lower panel), along with the bid-ask spreads ($\Delta$) and calibration mispricings ($E$). Mispricings exceeding bid-ask spreads are underlined. A value of zero means a value below one basis point. Index and tranches are quoted through the periodic premium, whereas the equity tranche is quoted as an upfront premium, see Section~\ref{sec:market}.}
\end{table}


We consider, as a first example, the calibration date May 13, 2005. We list in the upper panel of Table \ref{gpl:precrisis} the market data along with the calibration mispricings. Notice that all tranches are priced within the bid-ask spread. Then, as a second example, we switch to the calibration date October 11, 2005. We list in the lower panel of Table \ref{gpl:precrisis} the market data along with the calibration mispricings. This time some tranches are mispriced.

\begin{figure}[t]
\begin{center}
\includegraphics[width=\textwidth,trim=0 50 0 0]{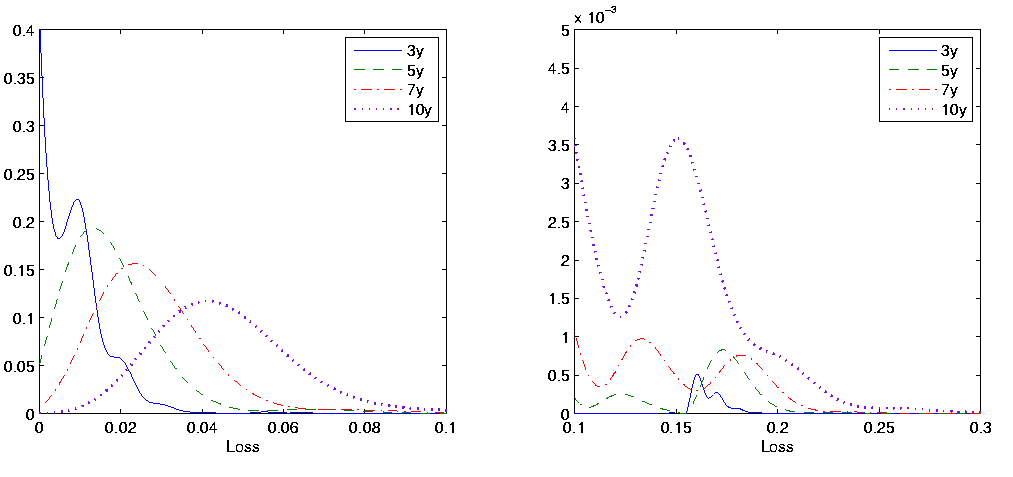}
\end{center}
\caption{
\label{fig:surfloss}
Loss distribution evolution of the GPL model with minimum jump size of $50$bp at all the quoted
maturities up to ten years, drawn as a continuous line.
}
\end{figure}

For all calibration dates we find that the loss distribution implied by the GPL model is multi-modal and the probability mass moves towards larger loss values as the maturity increases. These features are shared by different approaches. For instance, static models, such as the implied default-rate distribution  in Torresetti et al (2006c) suggest multi-modal loss distributions, as we have seen also here in our earlier sections on the implied copula. The evolution of the implied loss distribution is shown in Figure \ref{fig:surfloss}. Further, we notice that the values of the Poisson amplitudes are quite stable across the calibration dates. Indeed, in six months we observe at most four changes in their values, see  Brigo, Pallavicini and Torresetti (2006b).

\section{Application to more recent data and the crisis}\label{sec:recent}

In this section we check whether the critical features we discussed about implied correlation and the subsequent elements coming from more advanced models are still present in-crisis, after mid 2007. We will observe that the features are still present and often amplified in the market after the beginning of the crisis.

\subsection{Implied correlation in-crisis}

We begin by extending the sample of the Compound Correlation invertibility\index{Compound Correlation!invertibility} analysis to see if the non invertibility arose again during the market turmoil in 2007-2009. As before, we resort to the homogeneous pool model.

\begin{figure}[t]
\centering
\includegraphics[width=0.45\textwidth,clip]{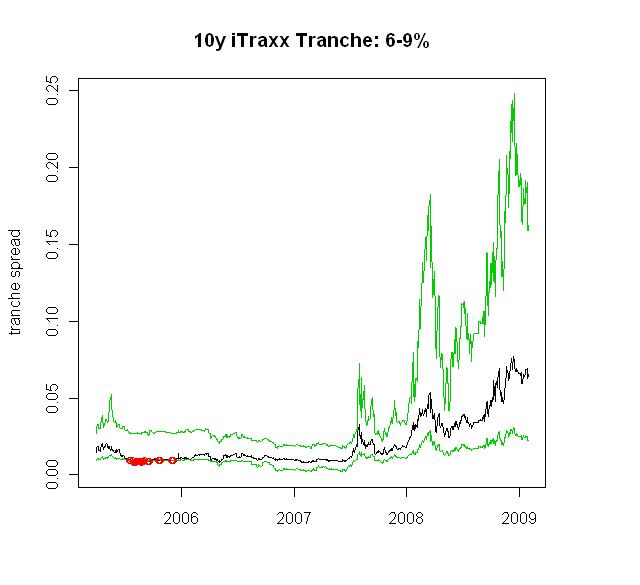}
\includegraphics[width=0.45\textwidth,clip]{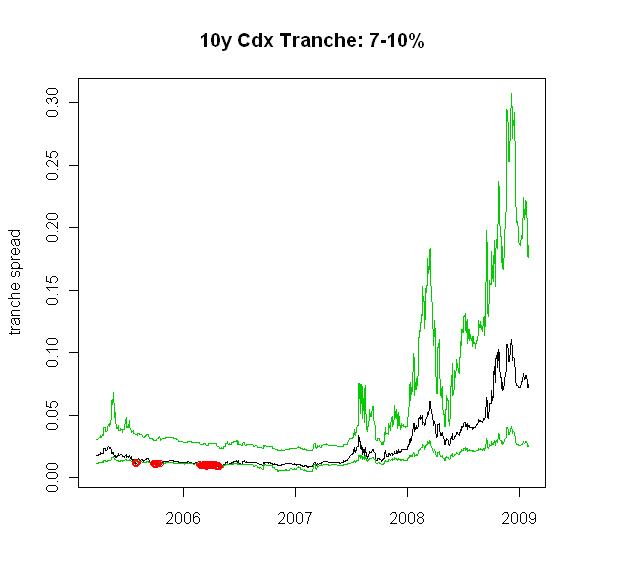}\\
\centering
\includegraphics[width=0.45\textwidth,clip]{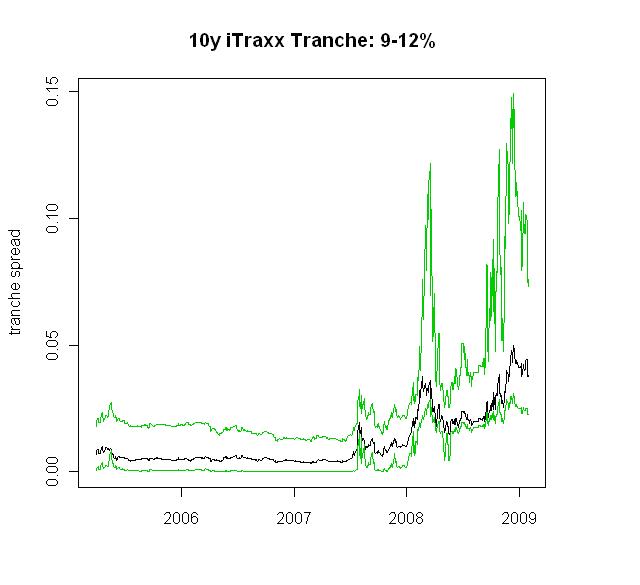}
\includegraphics[width=0.45\textwidth,clip]{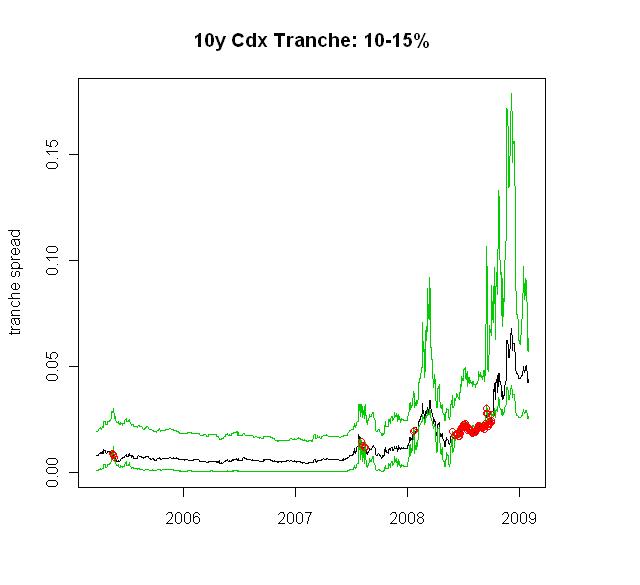}\\
\caption{
\label{fig:newInvCompItraxx5}
DJi-Traxx (left charts) and CDX (right charts) 10 year Compound Correlation Invertibility. Market spread (black line) versus minimum and maximum tranche spread obtained varying the compound correlation between 0 and 1 (green lines). Red dots are highlighting the dates and market spread that were not invertible (black line lies outside the area within the two green lines).
}
\end{figure}

In Figure~\ref{fig:newInvCompItraxx5} we show how since the credit crunch began in Summer 2007, the problematic tranches changed. In fact the non-invertibility interested mainly for the DJ-iTraxx the 10 year 12-22\% tranche, and for the CDX the 10 year 10-15\% and, marginally, the 10 year 15-30\% tranches (see Figure~\ref{fig:newInvCompItraxx5}).

Another relevant issue, besides non-invertibility of compound correlation, is its non-uniqueness when invertible. We have seen in Figure~\ref{fig:compCorrTrancheMappa} that on August 3, 2005 for some tranches the map between correlation and theoretical tranche spread is not monotonic.  In particular we have seen how this resulted to be the case for the DJi-Traxx 10 year 6-9\%tranche. Despite this map not being monotonic, on this particular date we did not encounter the problem of having more than one compound correlation implying the same theoretical spread.

%

\begin{figure}[t]
\centering
\includegraphics[width=0.45\textwidth,clip]{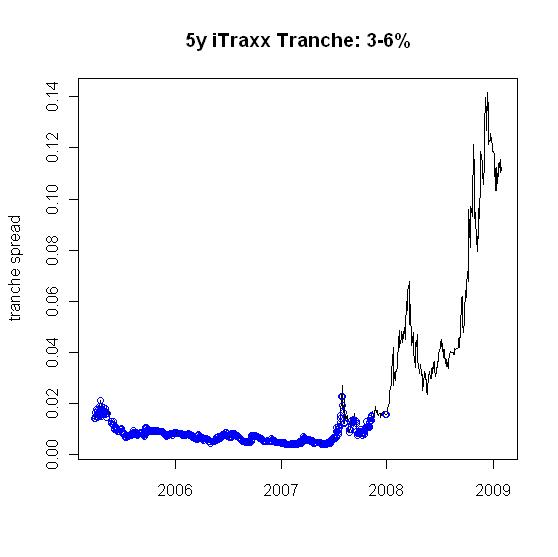}
\includegraphics[width=0.45\textwidth,clip]{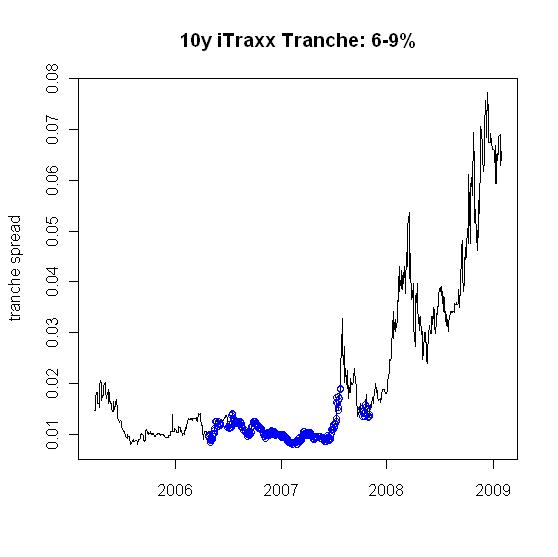}\\
\includegraphics[width=0.45\textwidth,clip]{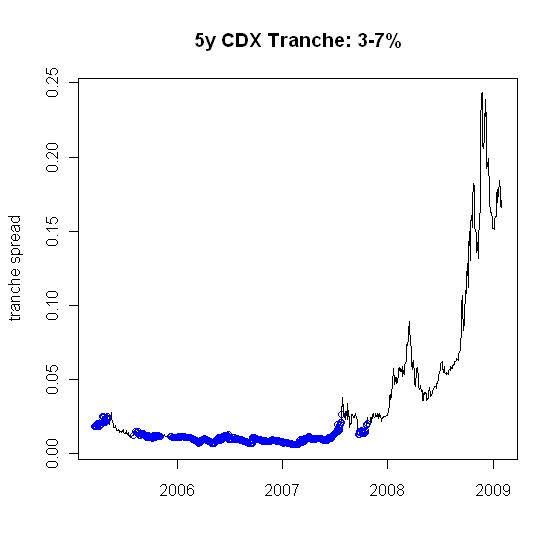}
\includegraphics[width=0.45\textwidth,clip]{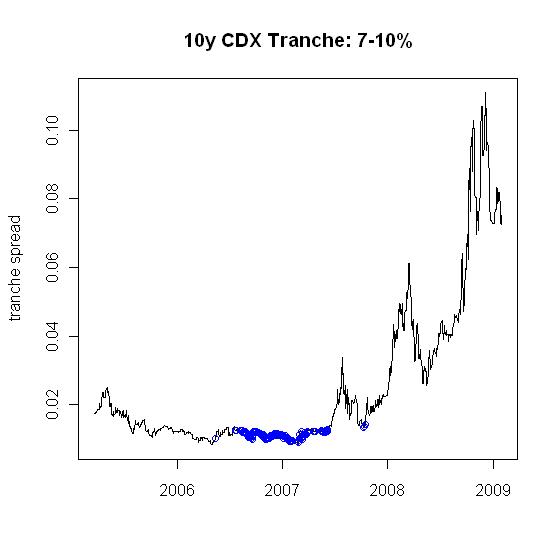}
\caption{
\label{fig:newUniCompItraxx}
Upper charts: DJi-Traxx 5 (left charts) and 10 (right charts) year Compound Correlation uniqueness. Lower charts: CDX. Blue dots are highlighting the dates where more than one compound correlation could reprice the tranche market spread.}
\end{figure}

We check whether there have been dates where a given tranche market spread could be mapped into more than one compound correlation (and we did a similar analysis for the CDX, not reported here).  In Figure~\ref{fig:newUniCompItraxx} we plot the market spread and highlight with dots the dates when the market spread was invertible into two different compound correlations for a few mezzanine tranches.

%

The problem of the non uniqueness of the compound correlation is particularly acute for several mezzanine tranches, some of which are shown in Figure~\ref{fig:newUniCompItraxx}, for both maturities, 5 and 10 years, and for both regions, DJi-Traxx and CDX.  Also for a given tranche the problem of non uniqueness of the compound correlation can span a relevant time window: for example the 5 year DJi-Traxx 3-6\% compound correlation has not been uniquely invertible from March 2005 to October 2007.

Now we move on to base correlation. Overall, it surpassed compound correlation and prevailed over more sophisticated dynamic loss models pre-crisis because:
\begin{itemize}
\item  it could be calibrated to almost all market conditions experienced pre-crisis;
\item  it resulted in a smooth map, thus allowing for pricing of bespoke tranches via interpolation;
\item  it could be used to define sketchy correlation mapping methods used for pricing tranches on Bespoke CDO portfolios, where no other correlations were available, see for example Bear Stearns and Lehman Brothers' reports by Reyfman et al. (2004) and Baheti et al. (2007);
\item  in its heterogeneous version it provided traders with easy to calculate hedge ratios.
\end{itemize}

In-crisis this perception was quite shaken.  For example the finite pool heterogeneous one-factor Gaussian copula base correlation often could not be calibrated to the 5 years CDX Series 9 tranches during 2008.


When the 15-30\% tranche spread cannot be calibrated this happens because the 30\% detachment base correlation cannot be increased enough: the minimum attainable theoretical 15-30\% tranche spread obtained for a 30\% detachment base correlation of 100\% is above the 15-30\% market tranche spread. This in turn occurs because the 15\% detachment base correlation is already so high that the 100\% upper bound for correlation values becomes a binding constraint.

\begin{figure}[t]
\centering
\includegraphics[width=0.65\textwidth,clip]{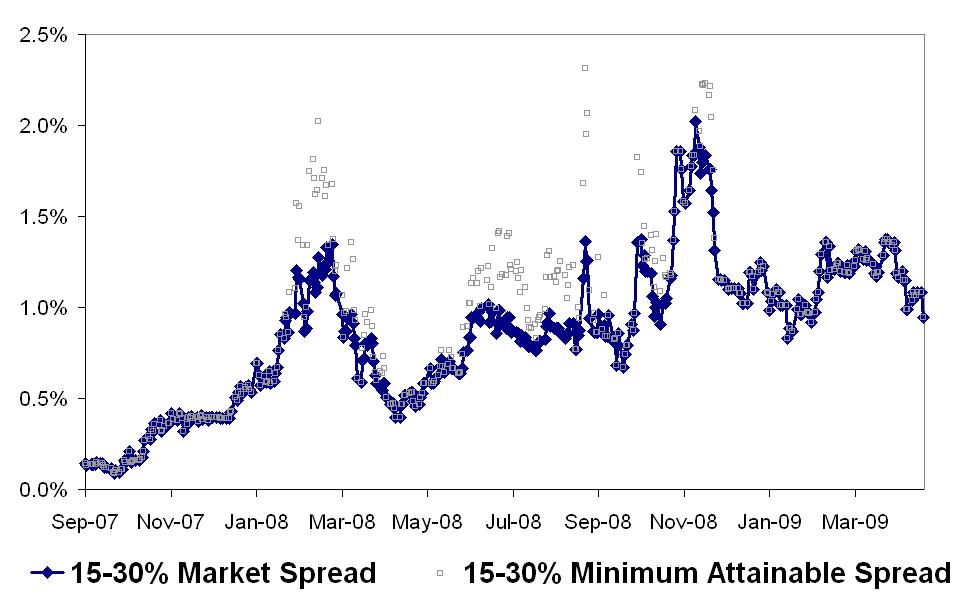}
\caption{
Market spread and minimum attainable spread via heterogeneous base correlation calibration for the 5 year CDX 15-30\% tranche.
\label{fig:minAttainAAA}
}
\end{figure}

In this respect we plot in Figure~\ref{fig:minAttainAAA} the CDX 5 year Series 9 15-30\% tranche market spread (blue line) and the minimum tranche spread (grey dots) that could be calibrated using the finite pool heterogeneous one-factor Gaussian copula base correlation from September 2007 to May 2009.

We have thus seen that the heterogeneous base correlation had calibration problems on those dates where the 100\% upper bound for the correlation became a binding constraint. If we move to an homogeneous assumption, the homogeneous copula could calibrate all the capital structure where the heterogeneous copula could not. In fact, allowing heterogeneity may change considerably the shape of the loss distributions.

\begin{figure}[t]
\centering
\includegraphics[width=0.65\textwidth,clip]{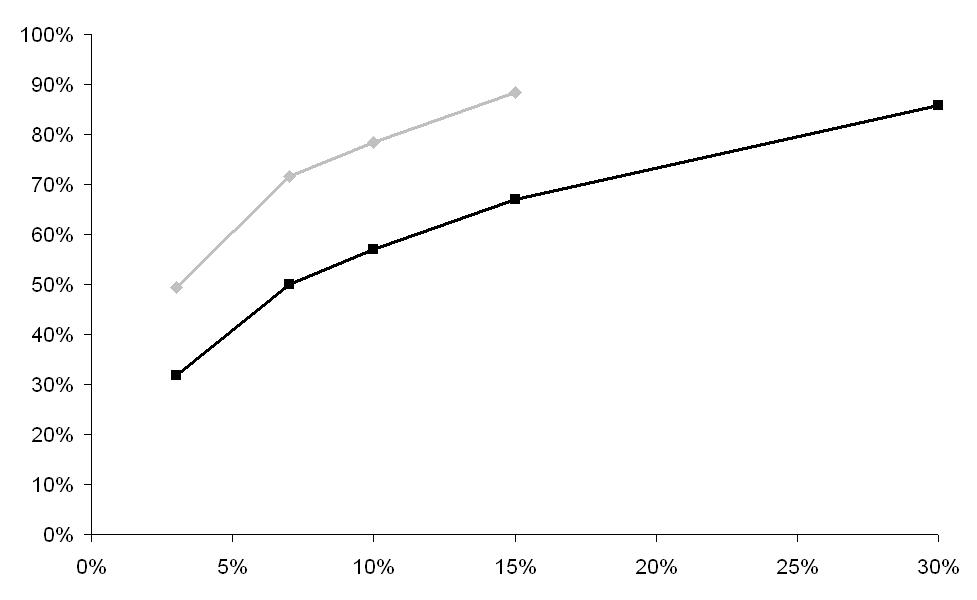}
\caption{
Heterogeneous (grey) and homogeneous (black) base correlation calibrated to the 5 year CDX tranche spreads on December 7, 2007.
\label{fig:baseHeterogHomog}
}
\end{figure}

In Figure~\ref{fig:baseHeterogHomog} we plot the 5 year CDX homogeneous and heterogeneous base correlation calibrated to the market spread on the 7th of December 2007.   We note that heterogeneous base correlation could not calibrate the 15-30\% tranche and we also note that the homogenous base correlation is smaller than the heterogenous base so that the 100\% upper bound for correlation was not reached.


\begin{remark}
We mention that the deterministic recovery assumption, whilst being computationally very convenient, does not help in the above setup. This has been addressed in the base correlation framework by Amraoui and Hitier (2008) and Krekel (2008), among others, who introduce random recovery. However, even with this update, base correlation remains flawed and may still lead to negative loss distributions, and anecdotally there are dates in-crisis where the heterogeneous pool base correlation with random recovery cannot fit the market if such random recovery is imposed to be consistent with single name CDS.
For a formulation of random recovery as a function of the systemic factor in the Impiled Copula framework see Remark \ref{rem:randrecimplcop}
\end{remark}

Summarizing, the heterogeneous version of the base correlation failed during the recent credit crisis as market participants have not been able to calibrate it for a considerable amount of time.

\subsection{Implied copula in-crisis}

Here we analyze the implied copula framework described in Section~\ref{sec:impliedcopula}, showing the implied loss distribution across time, so as to include the earlier examples as special cases of our analysis of the implied loss through history, both pre- and in-crisis.

\begin{figure}[ht!]
\begin{center}
\includegraphics[width=0.475\textwidth,clip]{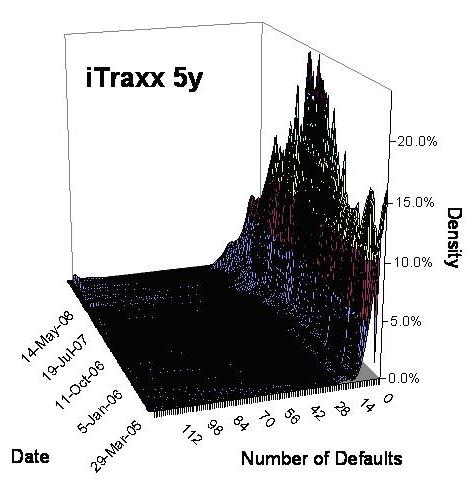}
\includegraphics[width=0.475\textwidth,clip]{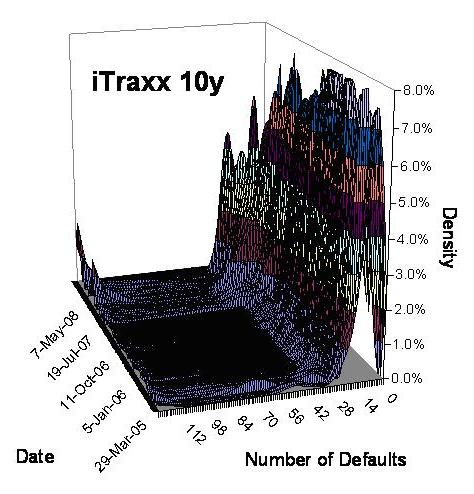}\\
\includegraphics[width=0.475\textwidth,clip]{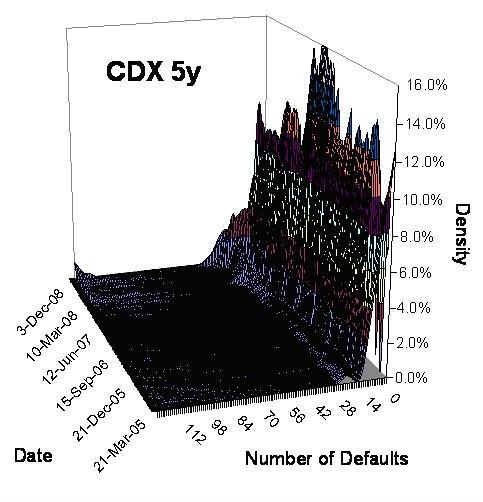}
\includegraphics[width=0.475\textwidth,clip]{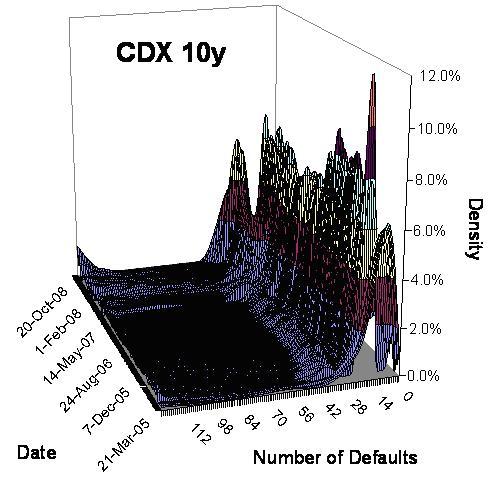}
\end{center}
\caption{ Implied distribution calibrated with the Implied Copula, via a two stage optimization as in Torresetti et al. (2006c), to the CDX and DJi-Traxx tranches for both the 5 and 10 years maturities calibrated from March 2005 to January 2009. See the full book by Brigo, Pallavicini and Torresetti (2010) for more details.  }\label{fig:implCopu_tri}
\end{figure}

In Figure~\ref{fig:implCopu_tri} we show the implied distribution for the default counting process, calibrated to the CDX and DJi-Traxx tranches (the entire capital structure except the super-senior tranche) and the index for both the 5 and 10 years maturities from March 2005 to January 2009.   We note how, since the beginning  of the crisis, the probability mass underlying the implied loss distribution shifted towards an higher number of defaults for both the DJi-Traxx and the CDX and for both maturities of 5 and 10 years. This can be interpreted as an increased perceived default riskiness in general.

Also, we note how since the start of the crisis the probability mass associated to a catastrophic or armageddon event, i.e. the default of the entire pool of credit references, has increased dramatically.   This is particularly visible for the 10 year default rate distributions for both DJi-Traxx and CDX. The increased probability of armageddon has been pointed out also in Morini and Brigo (2007, 2009) in the context of credit index options.

\begin{figure}[ht!]
\centering
\includegraphics[width=0.45\textwidth,clip]{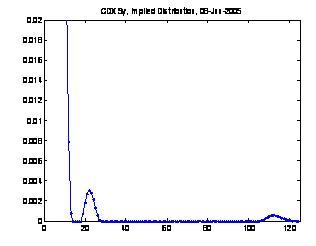}
\includegraphics[width=0.45\textwidth,clip]{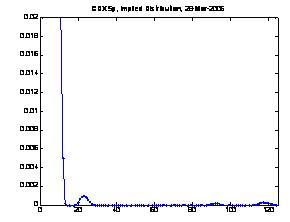}\\
\centering
\includegraphics[width=0.45\textwidth,clip]{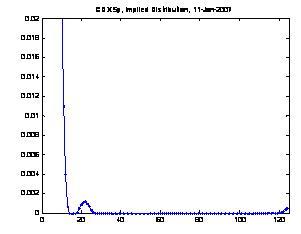}
\includegraphics[width=0.45\textwidth,clip]{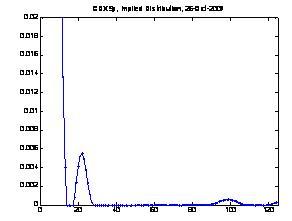}\\
\centering
\includegraphics[width=0.45\textwidth,clip]{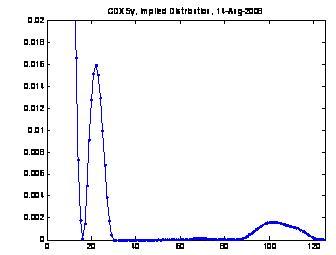}
\includegraphics[width=0.45\textwidth,clip]{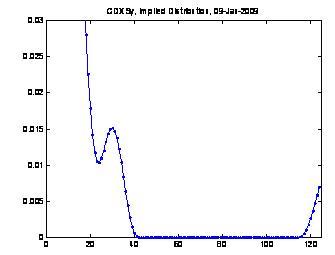}\\
\caption{ Implied distribution calibrated with the Implied Copula, via a two stage optimization as in Torresetti et al. (2006), to the CDX 5 year on-the-run tranches from March 2005 to January 2009.  }\label{fig:implCopu_hist1}
\end{figure}

In Figure~\ref{fig:implCopu_hist1} we show a feature of the implied distribution that cannot be immediately appreciated when inspecting Figure~\ref{fig:implCopu_tri}: the stability of the location of the modes in the implied loss distribution through time.

Taking a snapshot of the implied distribution calibrated to the CDX 5 year tranches, approximately every five months, in Figure~\ref{fig:implCopu_hist1} we note how the mode associated to about 20 names is quite constant in its location across all calibration dates.  In particular, on recent data, we note how even though the distribution shifts to the right, thus pricing an increased risk of default of the pool, the location of the mode does not seem to be affected.
In particular we examine data on March 2008 and January 2009: despite the massive shift to the right of the distribution, the non parametric optimization we implemented still shows a clear mode that would be associated to a cluster of defaults amounting to about 20 entities.
This observation leads almost naturally to the GPL dynamic framework outlined in Section~\ref{gplCrisis}.

%
%
%
%

\subsection{Expected Tranche Loss surface in-crisis}

The performance of the Expected Tranche Loss surface model in-crisis was not as egregious as pre-crisis. This was in part due to the fact that with the super-senior tranche\index{Super-Senior Tranche} available the system became over-determined.  In fact the capital structure was completed\footnote{Since the end of 2005 the liquidity of the super-senior tranche allowed us to include it in the set of the calibration instruments.} and the super-senior expected loss did not just have to fit the index spread, via the deterministic recovery, but had to retrieve both the index and the super-senior tranche.

Adding the super-senior tranche to the set of market instruments in the optimization of Equation~\eqref{eq:constrF} does not add any degrees of freedom in the variables whose function is being optimized, while adding two instruments: one for each maturity.

With the addition of the super-senior tranche the unknown ETL of the 22-100\% tranche for example will not be calibrated to the index assuming a deterministic recovery but will be calibrated instead directly to the super-senior tranche spread.  The index will then be a by-product of the calibration of all other tranches if one assumes a deterministic recovery.

We will modify the ETL approach to calibrate all tranches including the super-senior, thus spanning the entire capital structure, plus the index, by implying a deterministic recovery which is piecewise constant in time. We set
\[
(1 - R_t) d \defratenorm_t = d \lossnorm_t
\;;\quad
R_t := R_{5y}\, \forall t\le T_{5y} \;, R_t := R_{10y}\, \forall t> T_{5y}
\]
with $R_{5y}$ and $R_{10y}$ deterministic constants to be determined.

Furthermore, the objective function will be changed to penalize deviations of the step-wise constant calibrated recovery from 40\%, the market standard when pricing the recovery of Senior Unsecured CDS, leading to
%
%
\begin{eqnarray}
	\label{eq:constrFbis}
 	\underset{ \{ R_{5y},R_{10y},f(5y,3\%), ... ,f(10y,100\%) \} }{\operatorname{argmin}}
  	 100 \ \sum_{T}\sum_{A,B}{ ( \mbox{M{\tiny ISPR}S{\tiny TDZ}}^{A,B} ) ^2 }  +  \sum_{T} (R_T - 0.4)^2
\end{eqnarray}
subject to constraints~\eqref{eq:constrF2nd},
%
where the standardized mispricing is given by
\begin{equation*}
\mbox{M{\tiny ISPR}S{\tiny TDZ}}^{A,B} = \left\{
\begin{array}{ll}
 & ( S_0^{A,B,\mbox{theor}} - S_0^{A,B,\mbox{bid}} ) / S_0^{A,B,\mbox{ba}}  \ \ \ \mbox{if} \ \ \  S_0^{A,B,\mbox{theor}}  <  S_0^{A,B,\mbox{bid}}   \\
 & ( S_0^{A,B,\mbox{theor}} - S_0^{A,B,\mbox{ask}} ) / S_0^{A,B,\mbox{ba}}  \ \ \ \mbox{if} \ \ \  S_0^{A,B,\mbox{theor}}  >  S_0^{A,B,\mbox{ask}}   \\
 & 0  \ \ \ \mbox{otherwise}
\end{array}
\right.
\end{equation*}
with $S_0^{A,B,\mbox{ba}} := \onehalf \left( S_0^{A,B,\mbox{ask}} - S_0^{A,B,\mbox{bid}} \right)$.

More in particular on all dates where the optimization set out in~(\ref{eq:constrF}) results in a standardized mispricing of any tranche or index for any maturity larger than zero, i.e. the theoretical spread of any of the tranches or index for any of the two considered maturities is outside the bid-ask spread, we will run the optimization set out in~(\ref{eq:constrFbis}) where, with respect to the optimization~(\ref{eq:constrF}), we will introduce the piecewise constant recovery rate, flat between time 0 and the 5 year maturity and then between the 5 and 10 year maturity. This will give us two additional degrees of freedom in the optimization.

\begin{figure}[t]
\centering
\includegraphics[width=0.45\textwidth,clip]{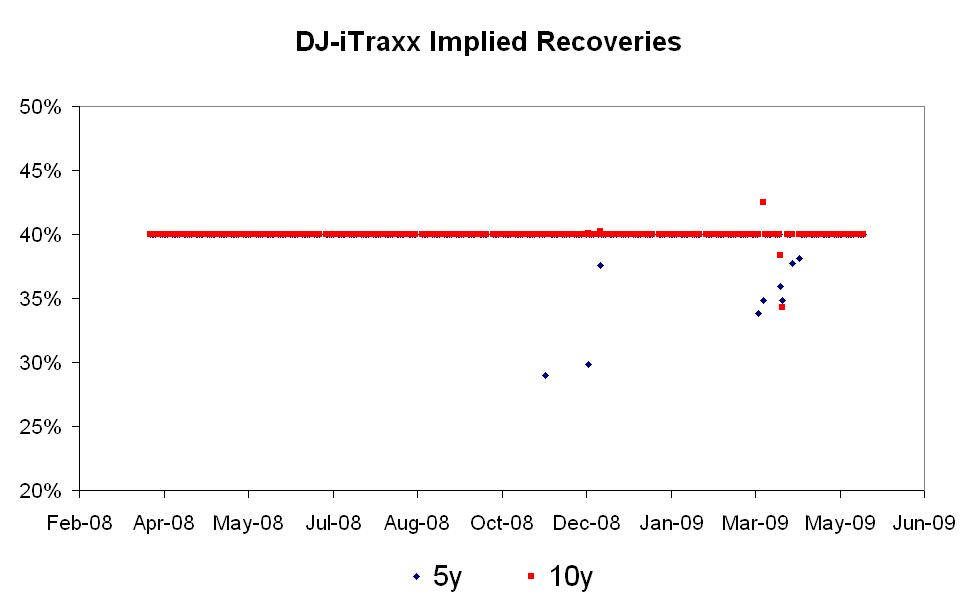}
\includegraphics[width=0.45\textwidth,clip]{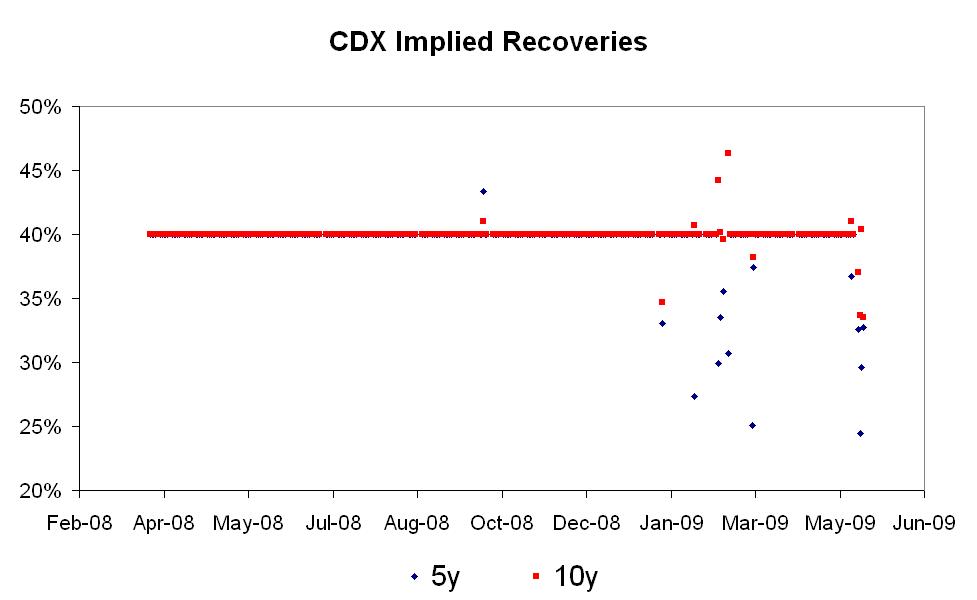}
\caption{ Stepwise constant Recovery Rates implied from tranches and index market quotes via the optimization set out in Equation~\eqref{eq:constrFbis}.  }\label{fig:ETL_impl_rec}
\end{figure}

In Figure~\ref{fig:ETL_impl_rec} we plot for each date the calibrated piecewise constant recovery: the 5 and 10 year implied recoveries.  We note that the above calibration has been implying a lower recovery than 40\% for the first 5 years, on those dates where a single fixed recovery was not able to calibrate all tranches and the index.    This is in line with the evidence that recovery rates depend on the business cycle, recovery rates being lower during periods of recessions. It is also consistent with the forecasts of a slower than expected economic recovery as we move out of the current crisis.\index{Recovery!business cycle}

%
%
%
%

%
%
%
%

\subsection{Consistent dynamic loss model in-crisis}\label{gplCrisis}

We resort now to the GPL model, to see how this model performs in crisis.
We change model formulation slightly, in order to have a model that is more in line with the current market, while maintaining all the essential features of the modeling approach. Compared with the model described in Section~\ref{sec:gplmodel}, we introduce the following modifications.
We fix the jump amplitudes $\alpha$ a priori. To do this, we reason as follows. Fix the independent Poisson jump amplitudes to the levels just above each tranche detachment, when considering a 40\% recovery.

For the DJi-Traxx, for example, this would be realized through jump amplitudes $a_i = \alpha_i /125$ where
%
\[
\alpha_5 = \mbox{roundup}\left(\frac{125 \cdot 0.03}{ (1-\rec)}\right),
\ \alpha_6 = \mbox{roundup}\left(\frac{125 \cdot 0.06}{ (1-\rec)}\right),
\ \alpha_7 = \mbox{roundup}\left(\frac{125 \cdot 0.09}{ (1-\rec)}\right),
\]
\[
\alpha_8 = \mbox{roundup}\left(\frac{125 \cdot 0.12}{ (1-\rec)}\right),
\ \alpha_9 = \mbox{roundup}\left(\frac{125 \cdot 0.22}{ (1-\rec)}\right),
\]
\[
\alpha_{10} = 125
\]
and, in order to have more granularity, we add the sizes 1,2,3,4:
\[
\alpha_1 = 1,
\ \alpha_2 = 2,
\ \alpha_3 = 3,
\ \alpha_4 = 4.
\]
In total we have $n=10$ jump amplitudes\footnote{We then modify slightly the obtained sizes in order to account also for CDX attachments that are slightly different.}. Eventually we obtain the set of amplitudes
\[
\alpha_i \equiv 125 \cdot a_i \in \{ 1, 2, 3, 4, 7, 13, 19, 25 , 46 , 125 \}
\]
Given these amplitudes, we obtain the default counting process fraction as
\[
\bar{C}_t = 1_{\{N_n(t)=0 \}} \bar{c}_t + 1_{\{N_n(t)> 0 \}}
\;,\quad
\bar{c}_t := \min\left( \sum_{i=1}^{n-1} a_i N_i(t) ,1\right) \ .
\]

Now let the random time $\hat{\tau}$ be defined as the first time where $\sum_{i=1}^{n} a_i N_i(t)$ reaches or exceeds the relative pool size of $1$.
\[
\hat{\tau} = \inf \{t: \sum_{i=1}^{n} a_i N_i(t) \ge 1   \} \ .
\]
We define the loss fraction as
\begin{eqnarray}
\bar{L}_t
:= 1_{\{\hat{\tau} > t \}} (1-\rec) 1_{\{N_n(t)=0 \}} \bar{c}_t
 + \,1_{\{\hat{\tau} \le t \}} \left[(1-\rec) 1_{\{N_n(\hat{\tau})=0 \}}  + 1_{\{N_n(\hat{\tau})> 0 \}} \left( 1 - \rec \, \bar{c}_{\hat{\tau}} \right)\right]\\ \nonumber
 = 1_{\{\hat{\tau} > t \}} (1-\rec) 1_{\{N_n(t)=0 \}} \bar{c}_t
 + \,1_{\{\hat{\tau} \le t \}}  \left( 1 - \rec \, \bar{c}_{\hat{\tau}} \right)
\end{eqnarray}

Notice that in this way whenever the armageddon component $N_n$ jumps the first time, the default counting process $\bar{C}_t$ jumps to the entire pool size and no more defaults are possible. Furthermore, whenever the armageddon component $N_n$ jumps the first time we will assume that the recovery rate associated to the remaining names defaulting in that instant will be zero. The pool loss however will not always jump to 1 as there is the possibility that one or more names already defaulted before the armageddon component $N_n$ jumped, and they defaulted with recovery $\rec$.
If at a given instant $t$ the whole pool defaulted, i.e. $\bar{C}_t=1$, this may have happened in two ways:

\begin{itemize}
\item $N_n$ jumped by $t$. In this case the portfolio has been wiped out with the contribution of an armageddon event. Notice that in this case $d\bar{C}_t=d\bar{L}_t$ if $N_n$ jumps at $t$.  In fact the recovery associated to the pool fraction defaulting in that instant will be equal to 0.

\item $N_n$ has not jumped by $t$. In this case the portfolio has been wiped out without the contribution of an armageddon event but because of defaults of more small or big sectors that do not comprise the whole pool.
Notice that in this case the loss is less than the whole notional, as all these defaults had recovery $\rec>0$.
\end{itemize}

In this way whenever $N_n$ jumps at a time when the pool has not been wiped out yet, we can rest assured that the pool loss will be above $1-\rec$. We do this because the market in 2008 has been quoting CDOs with prices assuming that the super-senior tranche would be impacted to a level impossible to reach with fixed recoveries at $40\%$.  For example there was a market for the DJi-Traxx 5 year $60-100\%$ tranche on 25-March-2008 quoting a running spread of 24bps bid.

We know how to calculate the distribution of both $\bar{C}_t$ and $\bar{L}_t$ given that:
\begin{itemize}
\item the distribution of $\bar{c}_t = \min\left( \sum_{i=1}^{n-1} a_i N_i(t) ,1\right)$ is obtained running a 'reduced' GPL, i.e. a GPL where the jump $N_n$ is excluded.
\item $N_n$ is independent from all other processes $N_i$ so that we can factor expectations when calculating the risk neutral discounted payoffs for tranches and indices.
\end{itemize}

Concerning recovery issues, in the dynamic loss model recovery can be made a function of the default rate $\bar{C}$ or other solutions are possible, see the recovery discussion in Section~\ref{sec:recoveryr} and also Brigo Pallavicini and Torresetti (2007) for more discussion. Here we use the above simple methodology to allow losses of the pool to penetrate beyond $(1-\rec)$ and thus affect severely even the most senior tranches, in line with market quotations.

Let us now assume we are to price two tranches: 2-4.8\% and 10-19.2\%.   The expected tranche loss of these tranches, and ultimately their fair spread, will depend primarily on the probability mass laying above the tranche detachment.

Bearing this interpretation of the modes in mind we decided to choose the GPL amplitudes by fixing the independent Poissons jump amplitudes to the level just above each tranche detachment considering a 40\% recovery. This led to the values $\alpha_5,\ldots,\alpha_{10}$ above for our $\alpha$s. With an eye to the richness of shapes of the Implied Copula historical calibrated distributions, we added four amplitudes (from 1 to 4), corresponding to a small number of defaults.

%
%

We now present the goodness of fit of the GPL just outlined through history. We measure the goodness of fit by calculating, for each date, the relative mispricing:
\[
\mbox{M{\tiny ISPR}R{\tiny EL}}^{A,B} = \left\{
\begin{aligned}
 & \frac{ S_0^{A,B,\mbox{theor}} - S_0^{A,B,\mbox{bid}} }{ S_0^{A,B,\mbox{mid}} }
   \ \ \ \mbox{if} \ \ \  S_0^{A,B,\mbox{theor}}  <  S_0^{A,B,\mbox{bid}} \\
 & \frac{ S_0^{A,B,\mbox{theor}} - S_0^{A,B,\mbox{ask}}}{ S_0^{A,B,\mbox{mid}} }
   \ \ \ \mbox{if} \ \ \  S_0^{A,B,\mbox{theor}}  >  S_0^{A,B,\mbox{ask}} \\
 & 0
   \ \ \ \mbox{otherwise}
\end{aligned}
\right.
\]
where $S_0^{A,B,\mbox{theor}}$ is the tranche theoretical spread as in equation (\ref{eq:tranche}), where in the calculation of the expecations of both numerator and denominator we take the loss distribution as resulting from the calibrated GPL.

\begin{figure}[t]
\centering
\includegraphics[width=0.45\textwidth,clip]{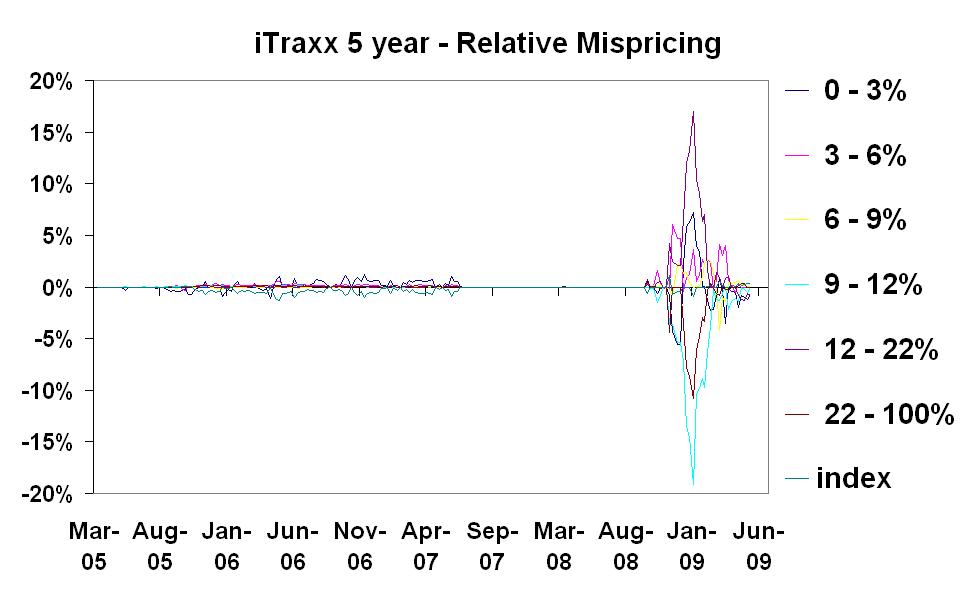}
\includegraphics[width=0.45\textwidth,clip]{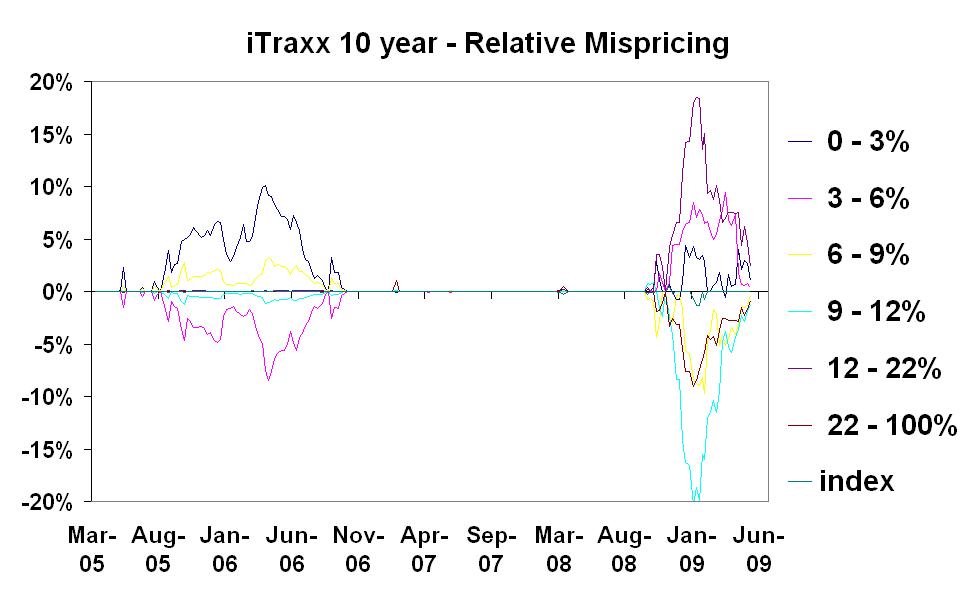} \\
%
%
\includegraphics[width=0.45\textwidth,clip]{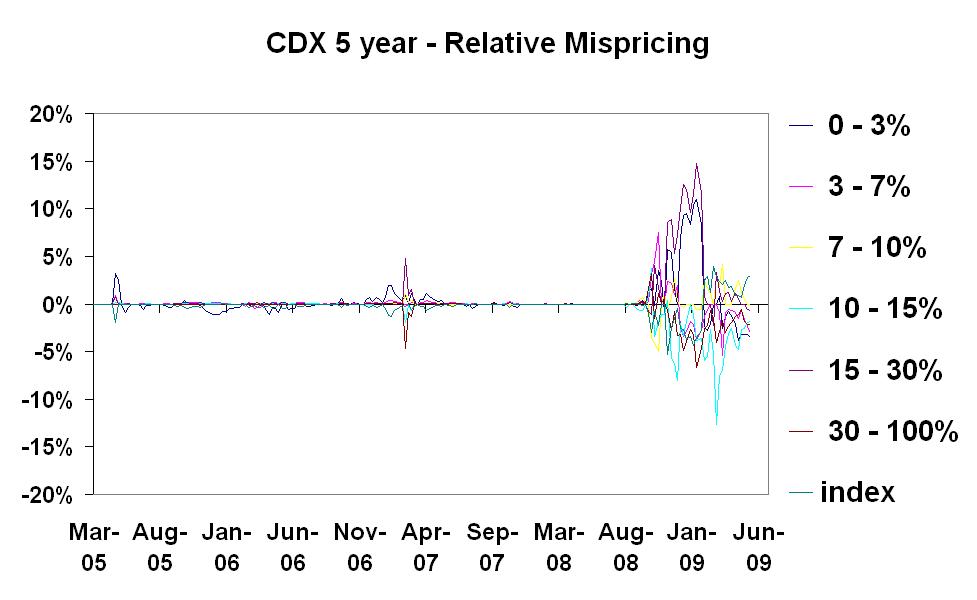}
\includegraphics[width=0.45\textwidth,clip]{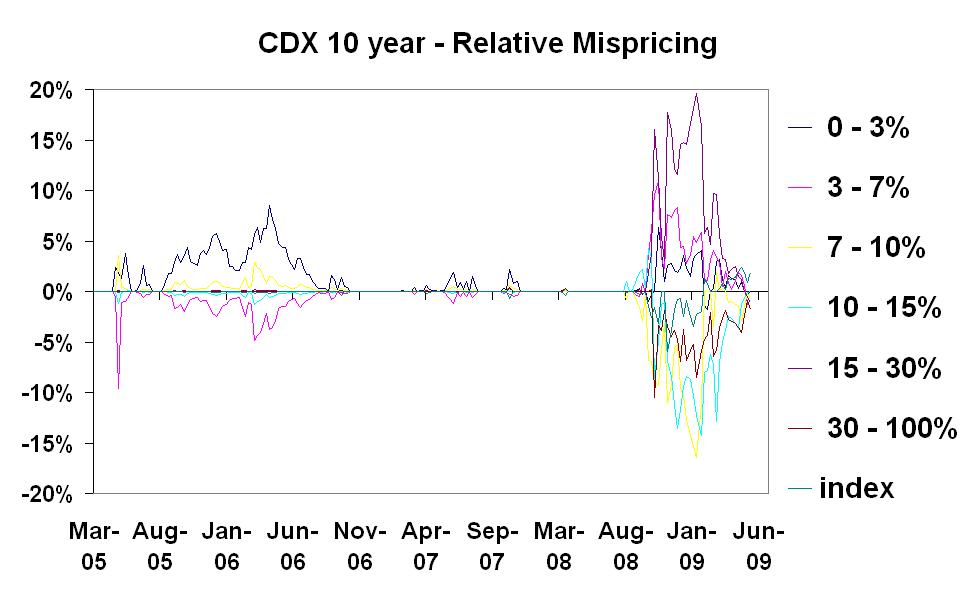}
\caption{ Relative mispricing resulting from the GPL calibration: DJi-Traxx (upper panels) and CDX (lower panels). }\label{fig:relMispr_GPL}
\end{figure}

In Figure~\ref{fig:relMispr_GPL} we present the relative mispricing for all tranches, for all maturities and for both indices throughout the sample: from March-2005 to June-2009. We note that while the 5 year tranches could be repriced fairly well in the current credit crisis for both DJi-Traxx and CDX, the 10 year tranches calibrations for both indices have been sensibly less precise  following the collapse of Lehman Brothers.  Recently, with the stabilization of credit markets we see  again fairly precise calibration results (i.e. a relative mispricing in the 2\% to 4\% range) for all tranches, both indices and both maturities.

To highlight where the problems in calibration come from, we have grouped mispricings across three different categories:
\begin{itemize}
\item  Instrument: we have grouped all tranches, independently of maturity and seniority, in one group and we have put the remaining calibrated instruments, i.e. the 5 year and 10 year indices, in the other group.
\item  Maturity: we have grouped all tranches and indices in two groups according to their maturity: 5 and 10 years.
\item  Seniority: we have grouped all tranches  (leaving out the indices) into three categories depending on the seniority of the tranche in the capital structure.
\begin{itemize}
\item  Equity: equity tranche for both the DJi-Traxx and CDX
\item  Mezzanine: comprising the two most junior tranches after the equity tranche.  For the DJi-Traxx this means the 3-6\% and 6-9\% tranches, whereas for the CDX this means the 3-7\% and the 7-10\% tranches.
\item  Senior: comprising the remaining most senior tranches. For the DJi-Traxx this means the 9-12\%, 12-22\% and 22-100\%  tranches, whereas for the CDX this means the 10-15\%, 15-30\% and 30-100\% tranches.
\end{itemize}
\end{itemize}

\begin{figure}[htp]
\centering
\includegraphics[width=0.7\textwidth,clip]{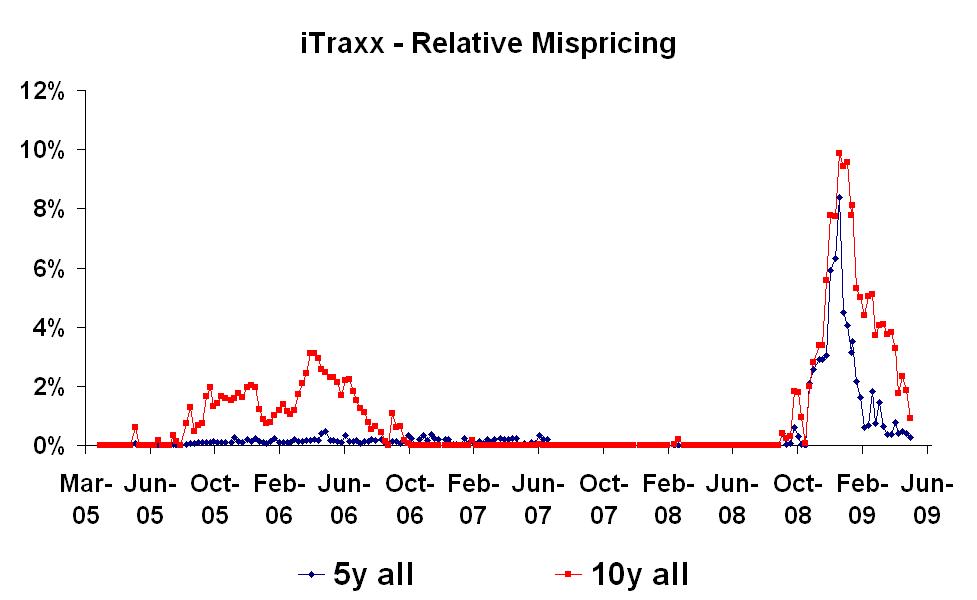} \\
\centering
\includegraphics[width=0.7\textwidth,clip]{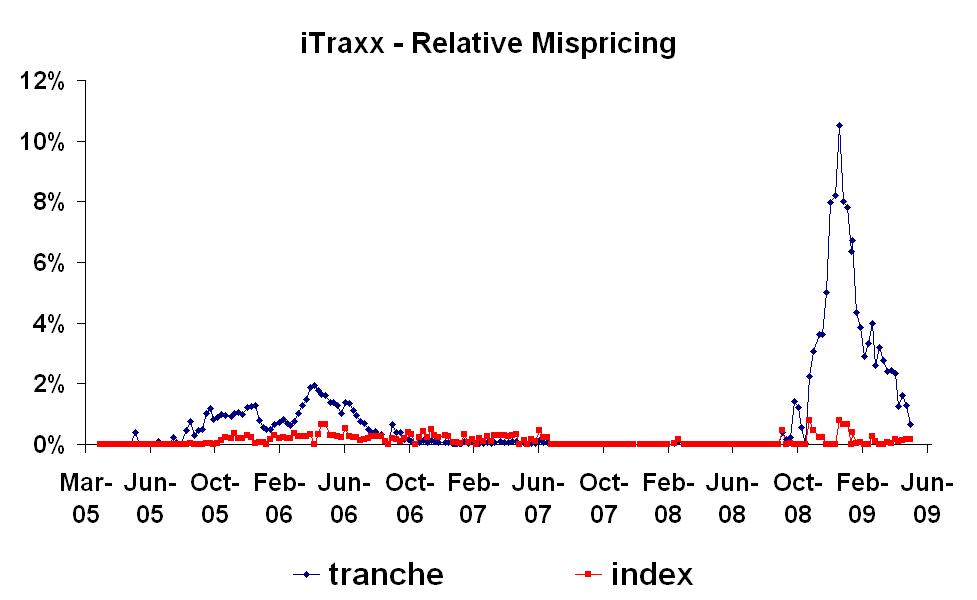}  \\
\centering
\includegraphics[width=0.7\textwidth,clip]{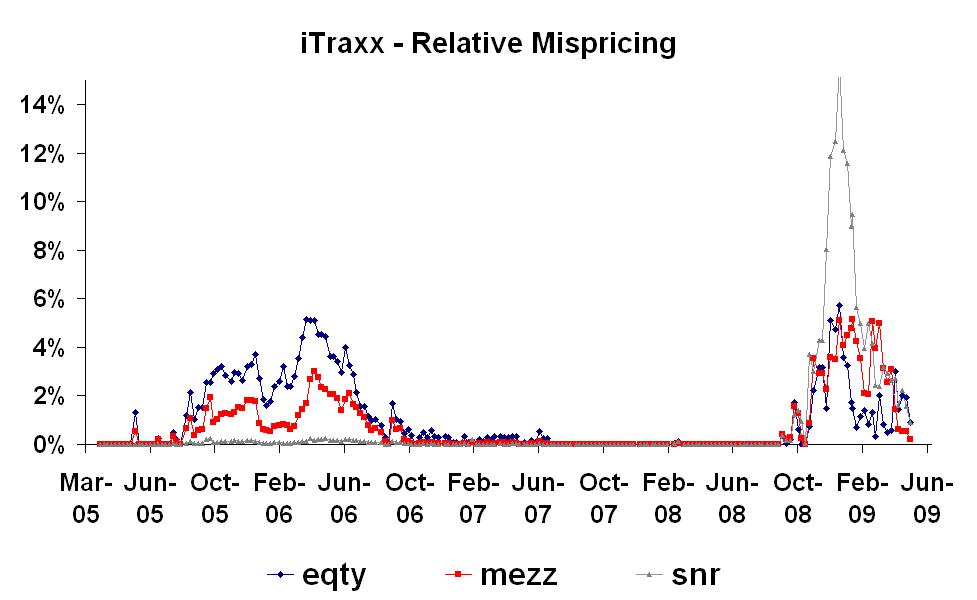}
\caption{ Relative mispricing resulted from the GPL calibration grouped by: Maturity (5 and 10 years), Instrument type (Index and Tranches) and Seniority (Equity, Mezzanine and Senior): DJi-Traxx. }\label{fig:itraxx_relMispr_GPL_groups}
\end{figure}

From Figure~\ref{fig:itraxx_relMispr_GPL_groups} (and its CDX analogous not reported here) we note that the GPL model produced calibrations that resulted in a relative mispricing that was larger (even though of a relatively small magnitude) in the period from June 2005 to October 2006.   For both the DJi-Traxx and the CDX the mispricing could be ascribed to the  10 year tranches, in particular the equity and mezzanine.

We also note that from October 2008 to June 2009 (in other words after the default of Lehman Brothers), the GPL calibration resulted again in a non zero but still fairly contained relative mispricing that in this case could be ascribed to both the 5 and 10 year tranches independently from their position in the capital strucuture.

%
%

\section{Final discussion and conclusions}\label{sec:conclusions}

We have followed a long path for Credit Derivatives and CDOs in particular, from the introduction of the Gaussian copula consistent at most with a single tranche at a single maturity, to the introduction of arbitrage-free dynamic loss models capable of calibrating all the tranches for all the maturities at the same time.

The critics we presented to the use of the Gaussian copula and of compound and base correlation had all been published before the beginning of the credit crisis. Thus, the notion that quantitative analysts and academics had no idea of the limits and dangers underlying the copula model is simply false. There is even a book titled ``Credit correlation: Life after copulas" edited by Lipton and Rennie (2007) that is the summary of talks given at a conference with the same name in 2006 in London.

Despite these warnings, the Gaussian copula model is still used in its base correlation formulation, although under some possible extensions such as random recovery. The reasons for this are complex. First the difficulty of all the loss models, improving the consistency issues, in accounting for single name data and to allow for single name sensitivities. While the aggregate loss is modeled so as to calibrate satisfactorily indices and tranches, the model does not see the single name defaults but just the loss dynamics as an aggregate object. Therefore partial hedges with respect to single names are not possible. As these are crucial in many situations, the market remains with base correlation.
Furthermore, even the few models that could try and have single name consistency have not been developed and tested enough to become operational on a trading floor. Indeed, a fully operational model with realistic run times and numerical stability is more than a prototype with some satisfactory properties that have been run in some ``off-line" studies. Also, when one model has been coded in the libraries of a bank, changing the model implies a long path involving a number of issues that have little to do with modeling and more to do with IT problems, integration with other systems, and the likes. Therefore, unless a new model appears to be greatly promising in all its aspects, there is reluctance in adopting it on the trading floor.

All these issues concurred in an unfinished research area. To this day there is no fully-tested operationally satisfying and single name consistent dynamic model capable of a consistent calibration of indices and CDO tranches across capital structure and maturity. It is true that to some extent this is an unfinished problem. This does not mean, however, that the quant community was unaware of the limitations of copula models before the crisis, as we abundantly documented.

Where does this leave us more generally? Modeling is just one of the many elements that may have contributed to the crisis, but certainly not the main one. As we broaden the picture, the number of factors that have little or nothing to do with modeling and that have been affecting the crisis becomes apparent. This helps us putting modeling into perspective as we close the paper.

There are several articles already that are good references on the crisis. We cite as a few examples quite different in nature Crouhy et al. (2008), Szego\footnote{Szego, G. (2009), The Crash Sonata in D Major. To appear in the Journal of Risk Management in Financial Institutions.} (2009), who presents an analysis of the crisis along a path starting with the investors ``feeding" on their houses, NINJAs\footnote{defined as ``An informal term for loans made in the USA to aspiring home-owners with 'No Income, No Job or Assets' (NINJA). More formally known as subprime mortgage lending, these loans, and the subsequent inability of many borrowers to repay them amid rising interest rates and falling house prices, were largely responsible for setting off a global credit crunch in mid-2007", The British Banker Association, see {\tt http://www.bba.org.uk/bba/jsp/polopoly.jsp?d=828\&a=10646}}, continuing with the frauds, partly documented by FBI\footnote{See for example the FBI Mortgage fraud report, 2007,\\ \url{www.fbi.gov/publications/fraud/mortgage$\_$fraud07.htm}.}, the problems of securitisation and of the originate-to-distribute system, the search for yield, the reversal in the real estate markets, the absent supervision, and finally the inadequate or wrong capital requirement rules and spillover effects. We also refer to El Namaki (2009a, 2009b) for an analysis of managers performances and quality before and during the crisis.

This is a good hint at illustrating the complexity of the crisis and the limited scope and roles mathematics and methodology have had in it. It should be clear by now that blaming mathematics for the sub prime fiasco, Lehman, Madoff, bail outs, unemployment and one of the worst recessions in recent history is missing the point.

We hope this paper has contributed to this understanding through its methodological journey and first hand witnessing of both the awareness some quants had of the models limitations and their attempts to surpass them. We also hope our forthcoming book, vastly expanding and updating this paper, will contribute to the debate. We believe the steps we - and many other researchers - took to address models limitations even before the crisis started are a good testimony of several Quants and Academics awareness and good will.

\newpage

\begin{center}
\noindent {\bf Table of Contents for the book}

 {\bf Credit Models and the Crisis: }

{\bf A journey into CDOs, Copulas,
Correlations and Dynamic Models }

{\bf Wiley, Chichester, 2010}
\end{center}

\contentsline{section}{Preface}{5}
\contentsline{section}{Part 1: Randomized by foolishness}{5}
\contentsline{section}{Part 2: How I learned to stop worrying and love the CDOs}{11}
\contentsline{section}{ \numberline{1}Introduction: credit modeling pre- and in-crisis}{13}
\contentsline{subsection}{ \numberline{1.1}Bottom-up models}{14}
\contentsline{subsection}{ \numberline{1.2}Compound correlation}{15}
\contentsline{subsection}{\numberline {1.3}Base correlation}{15}
\contentsline{subsection}{\numberline {1.4}Implied Copula}{16}
\contentsline {subsection}{\numberline {1.5}Expected Tranche Loss (ETL) Surface}{17}
\contentsline {subsection}{\numberline {1.6}Top (down) framework}{18}
\contentsline {subsection}{\numberline {1.7}GPL and GPCL models}{19}
\contentsline {subsection}{\numberline {1.8}Structure of the book}{20}
\contentsline {section}{\numberline {2}Market quotes}{23}
\contentsline {subsection}{\numberline {2.1}Credit indices}{23}
\contentsline {subsection}{\numberline {2.2}CDO tranches}{24}
\contentsline {section}{\numberline {3}Gaussian Copula model and implied correlation}{27}
\contentsline {subsection}{\numberline {3.1}One-factor Gaussian Copula model}{29}
\contentsline {subsubsection}{\numberline {3.1.1}Finite pool homogeneous one-factor Gaussian Copula model}{30}
\contentsline {subsubsection}{\numberline {3.1.2}Finite pool heterogeneous one-factor Gaussian Copula model}{31}
\contentsline {subsubsection}{\numberline {3.1.3}Large pool homogeneous one-factor Gaussian Copula model}{33}
\contentsline {subsection}{\numberline {3.2}Double-t Copula model}{35}
\contentsline {subsection}{\numberline {3.3}Compound correlation and Base correlation}{37}
\contentsline {subsection}{\numberline {3.4}Existence and non-monotonicity of market spread as a function of compound correlation}{38}
\contentsline {subsection}{\numberline {3.5}Invertibility limitations of compound correlation: pre-crisis}{40}
\contentsline {subsection}{\numberline {3.6}Base correlation}{41}
\contentsline {subsection}{\numberline {3.7}Is base correlation a solution to the problems of compound correlation?}{43}
\contentsline {subsection}{\numberline {3.8}Can the Double-t Copula flatten the Gaussian base correlation skew?}{45}
\contentsline {subsection}{\numberline {3.9}Summary on implied correlation}{46}
\contentsline {section}{\numberline {4}Consistency across capital structure: Implied Copula}{47}
\contentsline {subsection}{\numberline {4.1}Calibration of Implied Copula}{49}
\contentsline {subsection}{\numberline {4.2}Two stage regularization}{52}
\contentsline {subsection}{\numberline {4.3}Summary of considerations around Implied Copula}{54}
\contentsline {section}{\numberline {5}Consistency across capital structure and maturities: Expected Tranche Loss}{59}
\contentsline {subsection}{\numberline {5.1}Index and tranche NPV as a function of ETL}{60}
\contentsline {subsection}{\numberline {5.2}Numerical Results}{65}
\contentsline {subsection}{\numberline {5.3}Summary on Expected (Equity) Tranche Loss}{67}
\contentsline {section}{\numberline {6}A fully consistent dynamical model: Generalized-Poisson Loss model}{69}
\contentsline {subsection}{\numberline {6.1}Loss dynamics}{70}
\contentsline {subsection}{\numberline {6.2}Model limits}{71}
\contentsline {subsection}{\numberline {6.3}Model calibration}{71}
\contentsline {subsection}{\numberline {6.4}Detailed calibration procedure}{72}
\contentsline {subsection}{\numberline {6.5}Calibration results}{73}
\contentsline {section}{\numberline {7}Application to more recent data and the crisis}{77}
\contentsline {subsection}{\numberline {7.1}Compound Correlation in-crisis}{77}
\contentsline {subsection}{\numberline {7.2}Base Correlation in-crisis}{88}
\contentsline {subsection}{\numberline {7.3}Implied Copula in-crisis}{92}
\contentsline {subsection}{\numberline {7.4}Expected Tranche Loss surface in-crisis}{95}
\contentsline {subsubsection}{\numberline {7.4.1}Deterministic piecewise constant recovery rates}{98}
\contentsline {subsection}{\numberline {7.5}Generalized-Poisson Loss model in-crisis}{100}
\contentsline {section}{\numberline {8}Final discussion and conclusions}{111}
\contentsline {section}{Part 1: There are more things in heaven and earth, Horatio...}{111}
\contentsline {section}{Part 2: ... Than are dreamt of in your philosophy}{112}
\contentsline {section}{Notation and List of Symbols}{115}
\contentsline {section}{Index}{127}

\end{document}